\documentclass{article} 
\usepackage{colm2024_conference}
\usepackage{microtype}
\usepackage{url}
\usepackage{booktabs}
\usepackage{tcolorbox}
\tcbuselibrary{skins}
\usepackage{amsmath}
\usepackage{multirow}
\usepackage{balance}
\usepackage{tabularx}
\usepackage{subfigure}
\usepackage{graphicx}
\usepackage{fancyhdr}
\usepackage{lipsum} 
\usepackage{hyperref} 
\hypersetup{
    hidelinks
}
\colmfinalcopy

\fancypagestyle{page}{%
  \fancyhf{} 
  \renewcommand{\headrulewidth}{0.4pt} 
  
  \fancyhead[C]{LORE Technical Report}
  
  \fancyfoot[C]{}
}

\fancypagestyle{firstpage}{%
  \fancyhf{} 
  
  \renewcommand{\headrulewidth}{0.4pt} 
  \fancyhead[L]{\includegraphics[height=0.8cm]{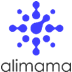}}
  \fancyhead[C]{LORE Technical Report}
  \fancyhead[R]{November 2, 2025}
  
  \renewcommand{\footrulewidth}{0pt} 
  \fancyfoot[C]{%
    \begin{minipage}{\textwidth}
    \footnotesize
    \rule{\textwidth}{0.4pt}\\[2pt]
    *Equal Contribution. Please refer to Sec~\ref{sec:contributor} for the complete list of contributors to this work.\\
    \end{minipage}
  }
}
  
\title{
\centering
LORE: A Large Generative Model for Search Relevance
}

\author{
  \begin{minipage}{\linewidth}
    \normalfont \centering
    Chenji Lu$^{*}$, Zhuo Chen$^{*}$, Hui Zhao, Zhiyuan Zeng, Gang Zhao, \\
    Junjie Ren, Ruicong Xu, Haoran Li, Songyan Liu, Pengjie Wang, Jian Xu, Bo Zheng \\[0.5em]
    \small \texttt{\{luchenji.lcj, cz462596, shuqian.zh, zengzhiyuan.zzy, zilong.zg,\\
    renjunjie.rjj, xuruicong.xrc, lhr476916, moxuan.lsy, pengjie.wpj, xiyu.xj\}@taobao.com} \\
    \small \texttt{bozheng@alibaba-inc.com} \\[0.5em]
    Search Advertising Team, Alimama (Alibaba Group)
  \end{minipage}
}

\newtcolorbox{mybox}[1]{
  colback=blue!10,        
  colbacktitle=blue!50,   
  coltitle=black,         
  title={#1},             
  fonttitle=\bfseries     
}
\newtcolorbox{greenbox}[1]{
  colback=green!10,        
  colbacktitle=green!50,   
  coltitle=black,          
  title={#1},              
  fonttitle=\bfseries      
}

\begin{document}
\thispagestyle{firstpage}
\maketitle

\begin{abstract}
\textbf{Achievement.} We introduce LORE(\textbf{\underline{L}}arge Generative M\textbf{\underline{o}}del for Search \textbf{\underline{R}}elevanc\textbf{\underline{e}}), a complete and sustainable framework of iterative practices for large language models(LLMs) in e-commerce search relevance, which achieves a cumulative +27\% improvement on the online GoodRate metric. Over the past three years, this project has demonstrated significant improvements in relevance judgment and has undergone three full-scale iterations across key dimensions, including data, features, training paradigms, evaluation, and application. Throughout the iterative development of LORE, we have gained valuable experience and insights that we believe are worth sharing with the community in this report.

\textbf{Insight.} To enhance LLMs for relevance, existing works have modeled the Chain-of-Thought (CoT) from various perspectives. However, we find that these methods often exhibit blind spots, as they lack a principled deconstruction of the task itself. Our analysis reveals that complex relevance judgment is not a monolithic reasoning problem but rather a composite of distinct capabilities, including knowledge and reasoning, multi-modal matching, and rule adherence. Based on this insight, we propose a systematic framework that first deconstructs the problem and then leverages this deconstruction to guide a training paradigm that explicitly models each required capability. We argue that such qualitative-driven analysis is essential for breaking through existing performance bottlenecks.

\textbf{Contributions.} LORE is a complete, replicable blueprint for LLM-based relevance modeling that spans the entire lifecycle.
First, we conducted systematic preliminary explorations into foundational training elements—including features, prompts, and base models—and summarized the general principles derived from this process.

Second, guided by our structural analysis, we propose a sophisticated two-stage training paradigm. In the first stage, we use progressive CoT synthesis and Supervised Fine-Tuning (SFT) to instill comprehensive capabilities. In the second, a carefully designed Reinforcement Learning (RL) phase aligns the model with human preferences. We also share key insights from our exploration of these training strategies.

Third, to ensure rigorous validation, we construct a comprehensive benchmark, RAIR, tailored to evaluate the core capabilities we identified.

Finally, to overcome the challenges of real-time computation, we designed a query-based stratified deployment strategy that comprehensively transfers the offline LLM's ability to online system, leading to substantial online performance gains. 

LORE serves as both a practical solution for developing advanced e-commerce relevance systems and a methodological reference for domain-specific post-training, with insights generalizable across vertical industries.

\end{abstract}
\newpage
\tableofcontents 
\newpage

\section{Introduction}
\vspace{-0.3cm}
Search relevance plays a pivotal role in e-commerce platforms such as JD and Taobao\citep{buy_2023}. The relevance model evaluates candidate items by determining their alignment with user queries, assigning relevance scores to filter out mismatched products\citep{buy3,whybuy}. This mechanism serves as a crucial foundation for enhancing user experience and search quality.

Large language models (LLMs) have exhibited remarkable capabilities and been extensively applied to relevance tasks. 
While state-of-the-art(SOTA) models \citep{deepseek,Qwen3,qwen2.5vl,gemini2.5} such as GPT-5 suffer from insufficient domain knowledge and prohibitive costs, post-training\citep{LPFT,walmartllm} smaller-scale LLMs on e-commerce data to develop domain-specific relevance experts has emerged as the predominant paradigm.

Recent advances\citep{walmartllm,LREF,JD2,TaoSR1} have progressively transitioned the post-training paradigm from rudimentary classification-oriented Supervised Fine-Tuning (SFT) to sophisticated enhancement of reasoning capabilities. While existing work has modeled the Chain-of-Thought (CoT)\citep{CoT} processes of LLMs from various perspectives to enhance their capabilities, significant blind spots remain. 
For example, ELLM\citep{ELLM} conceptualizes relevance as a process of attribute extraction and matching between queries and items. However, this approach falters in boundary cases that demand rule-based judgment. For instance, Lacking fine-grained rules for color discrimination, it might incorrectly match a "lake blue top" with a "sky blue top". To address this, subsequent approach like LREF\citep{LREF} and TaoSR1\citep{TaoSR1} advance beyond simple attribute matching by explicitly modeling rule-following CoT, thereby achieving rule-awareness. Nevertheless, their reliance solely on textual information introduces a significant limitation: the inability to process visual cues. As a result, given the query "blue top," the models cannot confirm a match if an item's title omits the color, even its image clearly displays a blue top. 
These limitations ultimately stem from a failure to address a fundamental question that defines the capabilities required for relevance assessment: \textit{What constitutes a comprehensive CoT representation for such tasks?}

\vspace{-0.4cm}
\begin{figure}[!hb]
  \centering
\includegraphics[width=0.95\textwidth]{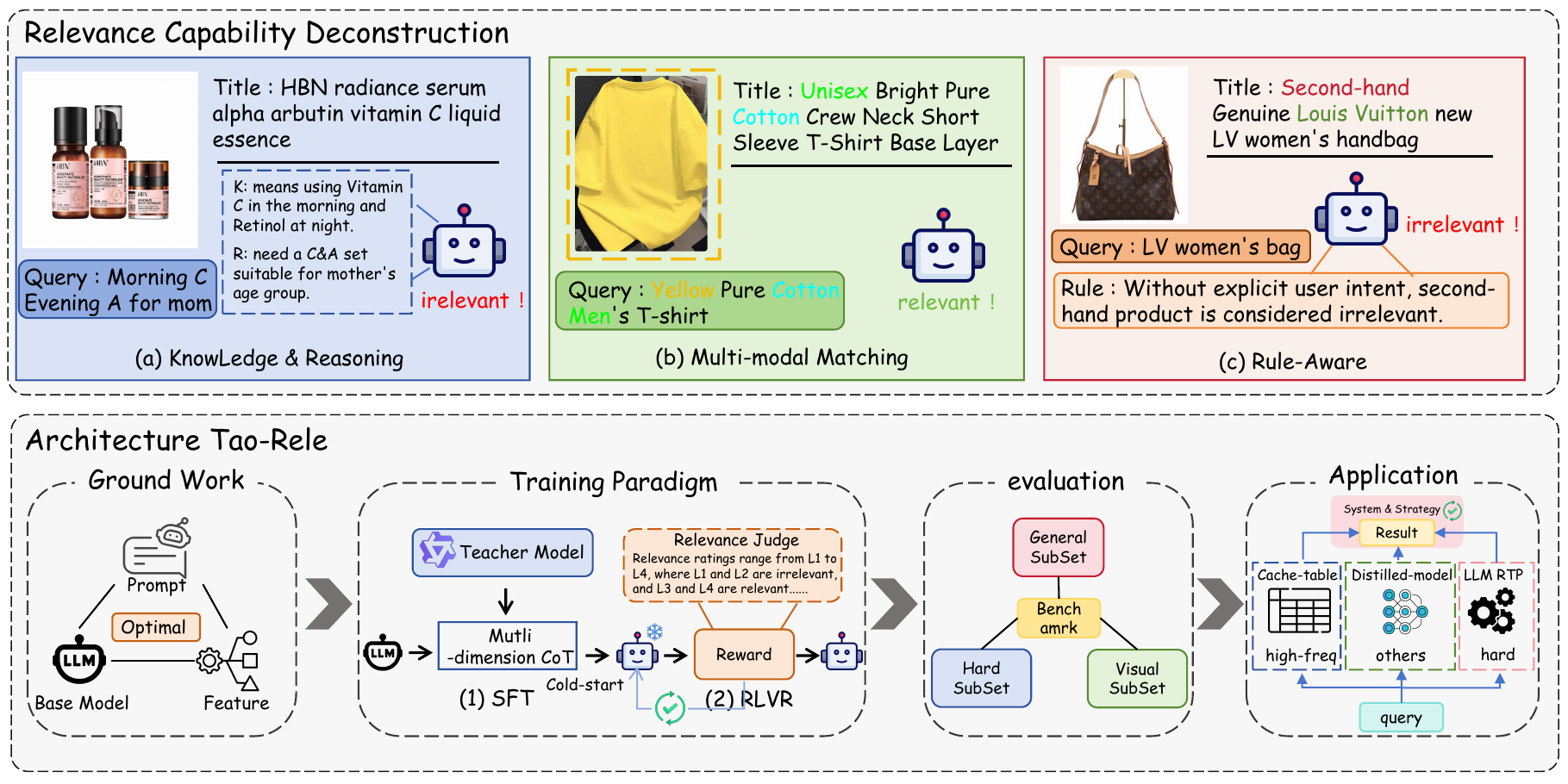}
  \vspace{-0.4cm}
  \caption{Overview of the LORE theoretical framework and architecture.}
  \vspace{-0.4cm}
  \label{fig:intro}
\end{figure}

To systematically address this question, we first deconstruct the relevance assessment, as detailed in Section~\ref{prel}. Our analysis breaks this process down into two fundamental steps: (1) \textbf{Path Construction:} mapping query requirements to specific attributes and establishing semantic paths between queries and products; (2) \textbf{Path Following:} executing precise judgments along this path on an attribute-by-attribute basis. This two-stage decomposition allows us to pinpoint the essential capabilities that model must possess, which we summarize as follows:
(1) Knowledge and Reasoning\citep{deepseek,skywork,kimi1.5,knowledge1,knowledge2}. Knowledge is required to decipher domain-specific terminology in queries (e.g., "morning C, evening A" as Vitamin C and retinol) and interpret ambiguous item details. Reasoning then leverages this knowledge to infer higher-level user intent (e.g., deducing that "for mom" implies a middle-aged demographic), as illustrated in Fig.~\ref{fig:intro}(a).
(2) Multi-modal Matching\citep{multi-modal1,multi-modal2,multi-modal3}. Since item attributes are often distributed across various modalities, this ability is crucial. The model must perform cross-modal alignment, connecting textual query requirements with visual evidence from the item. As shown in Fig.~\ref{fig:intro}(b), a query for "yellow" must be validated against the item's image when the color is not mentioned in its textual description.
(3) Rule-Aware. Rules serve as the concrete embodiment of human subjective judgment, supplementing general logic by providing explicit, quantifiable, and reproducible criteria for relevance assessment. it must adhere to these predefined rules to make nuanced judgments. For instance, as depicted in Fig.~\ref{fig:intro}(c), classifying a "second-hand LV bag" against a query for "LV bags" is not merely an attribute-matching problem. It requires applying a specific rule about item condition (e.g., new vs. used) to ensure accuracy.

To instill these capabilities, we introduce LORE 3.0(\textbf{\underline{L}}arge Generative M\textbf{\underline{o}}del for Search \textbf{\underline{R}}elevanc\textbf{\underline{e}}), our most recent iteration of the systematic LLM practice framework for relevance task, as depicted in the lower part of Fig~\ref{fig:intro}. 
Firstly,  we establish a robust foundation by conducting preliminary explorations to identify the optimal base model, feature set, and prompt structure. 
Secondly, building on this foundation, we propose a two-stage training paradigm encompassing SFT and Reinforcement Learning(RL). The SFT stage is designed to instill comprehensive reasoning abilities. Following our two-step discrimination framework, we synthesize reasoning trajectories through a progressive synthesis strategy and elevate the model's capability upper bound via distillation. The subsequent RL stage aims to align the model with human preferences. To this end, we adapt Reinforcement Learning with Verifiable Rewards (RLVR)\citep{deepseek} and tailor it to the nuances of relevance tasks. This process prunes flawed reasoning paths while preserving the rich, multi-dimensional capabilities established during SFT.
Thirdly, to rigorously evaluate our model, we constructed a comprehensive and challenging benchmark designed specifically around the core capabilities identified earlier. This allows for a systematic assessment of the model's performance across each dimension.
Finally, we implemented a practical strategy to transfer these enhanced LLM capabilities into our online system. By stratifying queries and applying differentiated deployment policies, we achieved a cumulative 27\% improvement in GoodRate. In summary, the main contributions of this work are as follows:

\begin{itemize}
    \item We present a theoretical deconstruction of the relevance task, identifying a core set of model capabilities. This analysis establishes a principled blueprint for designing future relevance models.
 \item Building on the theory, we introduce LORE, a replicable, end-to-end methodology for the entire model lifecycle—from data synthesis and training to online deployment. It serves as a comprehensive guide for industrial practice.
 \item We demonstrate LORE's effectiveness through comprehensive offline and online validation. The framework not only achieves SOTA performance on offline benchmarks but also delivers a substantial 27\% cumulative improvement in online GoodRate upon deployment.
\item We distill key insights and best practices from our extensive exploration of SFT and RL strategies. These findings provide actionable guidance for fine-tuning LLMs in other specialized domains:
\end{itemize}

\vspace{-0.3cm}
\begin{mybox}{Key Findings during SFT}
\textbf{1. More features beat fewer.} Providing stable and relevant information, even if redundant, can yield gains for the model(Section~\ref{sec:3.2.1}).

\textbf{2. A concise prompt with essential information works best. }Information-rich but verbose prompts do not necessarily yield improvements, whereas prompts with insufficient key information lead to degradation in model capability(Section~\ref{sec:3.2.3}).

\textbf{3. Data scaling law shows diminishing returns.} Model performance grows rapidly initially but plateaus as data increases.
(Section~\ref{sec:3.3.3}).

\textbf{4. Naive teacher CoT distillation results in negative effects.} Direct teacher CoT distillation underperforms vanilla SFT due to distribution shift.
(Section~\ref{sec:6.1})
\end{mybox}

\begin{greenbox}{Key Findings during RL}
\textbf{1. Curriculum learning still works.} Curriculum learning strategies based on difficulty grading significantly outperform randomly ordered strategies.(Section~\ref{sec:3.4.3})

\textbf{2. Entropy collapse rate caps model performance.} In the early stages of training, the model rapidly trades entropy for performance gains. When entropy decreases to a certain level, the model's exploration capability becomes constrained, and subsequent room for improvement diminishes.(Section~\ref{sec:3.4.3})

\textbf{3. Smart strategies effectively slow entropy collapse.} The report discusses several optimization methods including clip-higher, on-policy, and entropy loss, ultimately finding that clip-higher achieves optimal performance.(Section~\ref{sec:3.4.3})

\textbf{2. Long CoT is not necessary for better performance.} No significant increase in model output length was observed during the RL process, indicating that long CoT are merely a potential byproduct when the model's capability improves.(Section~\ref{sec:6.2})

\end{greenbox}

\section{Theoretical Framework}
\label{prel}
In this section, our goal is to develop a comprehensive view of the relevance task and, on this basis, to identify the key modeling capabilities required for effective e-commerce search. These capabilities will serve as the foundation for the methods introduced in the subsequent sections.

\subsection{Task Definition}
\begin{figure}[htbp]
  \centering
  \includegraphics[width=0.95\textwidth]{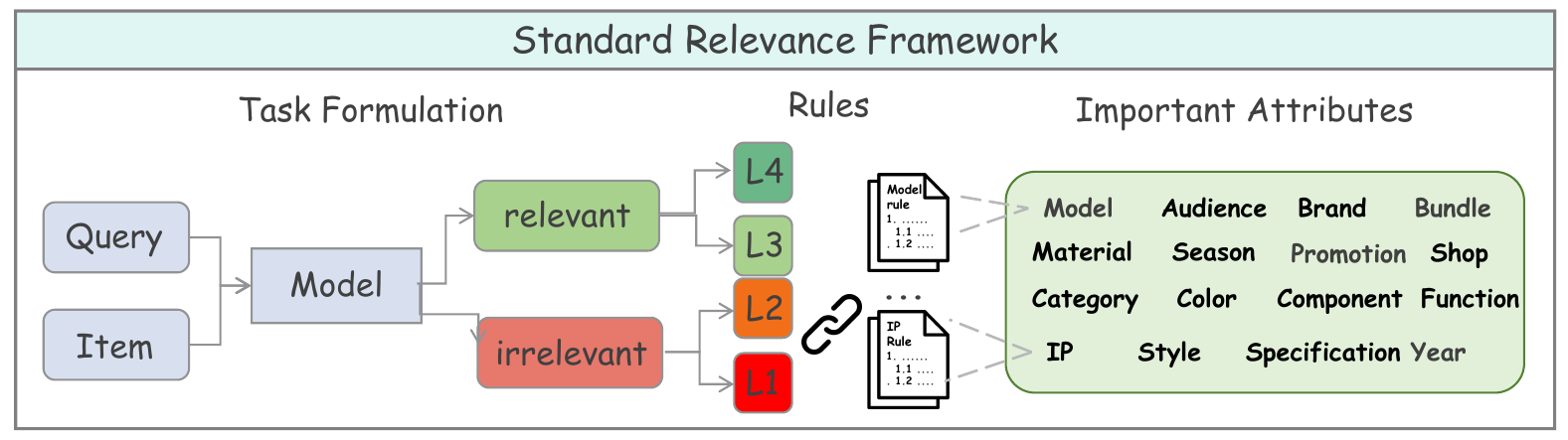}
  \caption{A standard definition of the relevance task}
    \label{fig:framework}
\end{figure}

\subsubsection{Problem Formulation}
At its core, e-commerce search can be conceptualized as a mapping from user queries to product items. The role of relevance modeling is to refine this mapping by establishing a quantitative measure—relevance—that evaluates the quality of this match. This measure accounts for both objective criteria (e.g., attribute and constraint satisfaction) and subjective factors (e.g., user intent and preference).
Formally, as illustrated in Fig~\ref{fig:framework}, given a query $Q_i$ and an item $I_i$, the model performs pointwise discrimination to determine whether the pair is relevant or irrelevant. In its simplest form, this can be viewed as a binary classification problem.

\subsubsection{Fine-Grained Relevance Judgment}
A purely binary formulation, however, cannot adequately characterize the varying degrees to which $I_i$ satisfies the information need expressed by $Q_i$. To address this limitation, we adopt a more fine-grained discrimination framework with four levels, denoted L1–L4, where higher levels indicate a greater degree of satisfaction of $Q_i$ by $I_i$. Concretely, L1 and L2 are treated as \emph{irrelevant}, while L3 and L4 are treated as \emph{relevant}. This graded scheme enables the model to capture more nuanced distinctions in relevance beyond a simple yes/no decision.

\subsubsection{Attribute Schema for Query Demands}
Orthogonally to the four-level relevance scale, we decompose the potential demands encoded in queries into 18 attribute dimensions (e.g., category, brand, style, and others). For each query–item pair, we assess the extent to which the item satisfies the attributes explicitly or implicitly expressed in the query, and use these attribute-level judgments to ground and justify the assigned relevance level. This attribute schema provides a structured lens through which relevance decisions can be interpreted and analyzed.

\subsection{Deep Analysis of Relevance Modeling}

To systematically quantify relevance within the e-commerce search context, we decompose the assessment process into three synergistic components: \emph{item understanding}, which constructs the attribute space; \emph{query understanding}, which decodes user intent into retrieval pathways; and the subsequent \emph{relevance judgment}.

\subsubsection{Item Understanding}
In relevance judgment, item understanding can be analyzed along two main dimensions: (i) the nature of item attributes and (ii) the presentation modality.
\begin{figure}[htbp]
    \centering
    \includegraphics[width=1\linewidth]{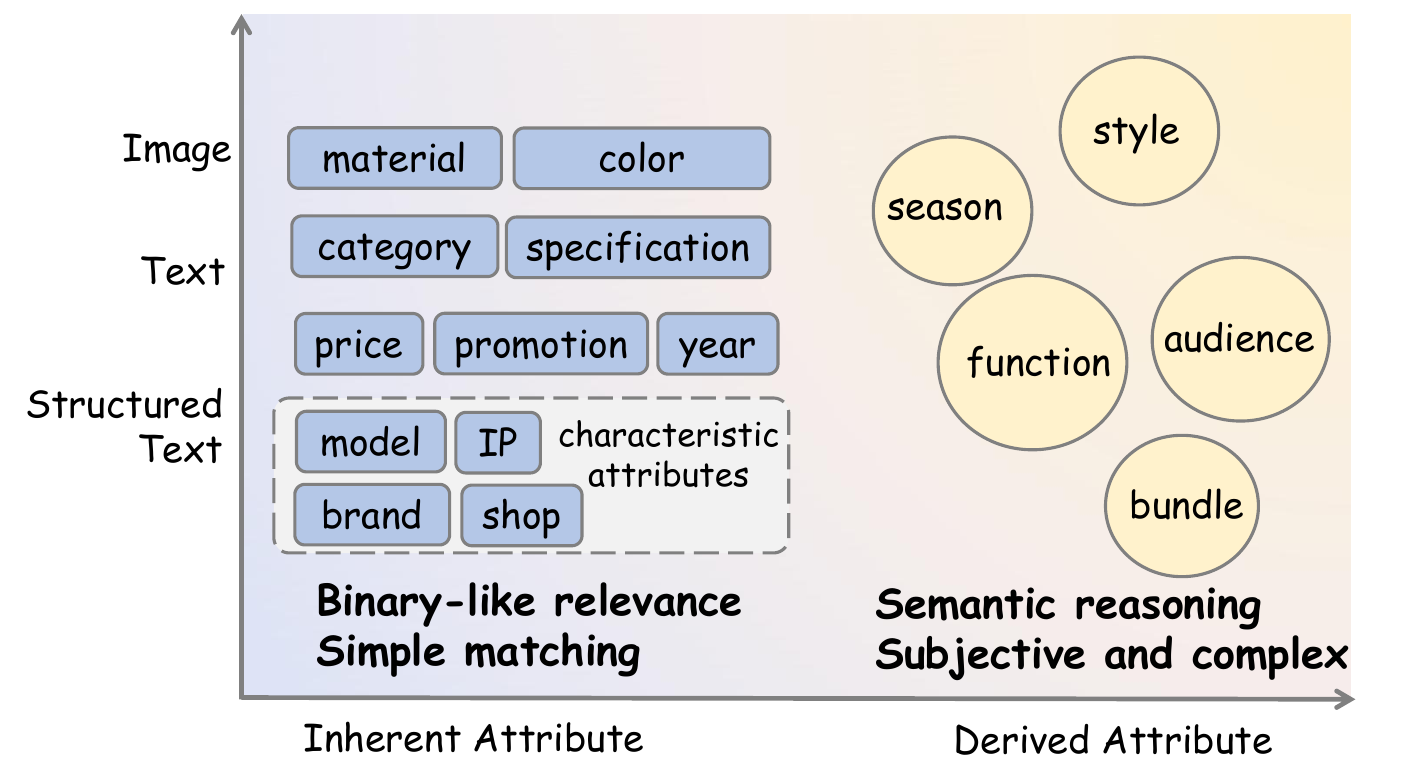}
    \caption{Comprehensive Overview of Item Attributes}
    \label{fig:placeholder}
\end{figure}

\paragraph{Attribute Properties}
From the perspective of attribute properties, we distinguish between \emph{inherent attributes} and \emph{derived attributes}. The key difference lies in whether relevance judgment can be made with direct observation and simple matching, or whether it requires additional reasoning and stronger semantic understanding.
\begin{itemize}
    
\item  \textbf{Inherent attributes} denote objectively existing and reliably observable properties of the item. They are usually direct consequences of the product design or the physical carrier, such as category, shop, brand, IP, model, specification, material, color, price, year, and promotion. For such attributes, relevance tends to be close to a binary decision (relevant vs.\ not relevant), and the decision rules are relatively clear. Within inherent attributes, we can further distinguish a subset of \emph{characteristic attributes}, which are the most central and discriminative attributes, e.g., brand, model, IP, and shop name. On e-commerce platforms, these attributes are typically expressed in a clear, objective, and relatively standardized form. Consequently, for query--item matching that includes characteristic attributes, simple term-matching strategies often suffice for effective relevance judgment.

\item  \textbf{Derived attributes} are induced from inherent attributes and usually correspond to product goals or usage effects. These attributes are not always directly observable, and relevance judgment relies more heavily on semantic understanding and reasoning, often involving subjectivity and degrees of satisfaction (e.g., ``style'', `` audience'', ``function''). This makes the decision problem substantially more complex. A major challenge is that different industries and categories develop their own attribute systems and surface forms, which are highly diverse and difficult to enumerate exhaustively.

\end{itemize}

\paragraph{Presentation Modality}
Presentation modality describes how attributes are exposed to the user. From this perspective, item information can be broadly divided into text-only and multi-modal (text+image) representations. The modality determines whether item understanding must rely on multi-modal modeling, and it interacts with the nature of attributes. For some attributes (e.g., color, style, material), image information is often indispensable; for others (e.g., brand, price, year), text is usually sufficient for accurate identification and judgment. In multi-modal scenarios, different attributes depend on visual information to different degrees. When characteristic attributes are expressed consistently in both text and images, multi-modal capabilities can further improve robustness. For highly subjective derived attributes (e.g., style, effect), joint understanding of text and images becomes critical for high-quality relevance judgment.

Overall, at the item-understanding level, relevance judgment for inherent attributes mainly depends on accurate query understanding and e-commerce domain knowledge, and is relatively direct. In contrast, derived attributes pose higher difficulty both in path construction (reasoning) and in relevance judgment (semantic understanding).

\subsubsection{Query Understanding}

In contrast to the structured attributes of items, user queries are often free-form and express a diverse range of needs. The primary challenge in Query Understanding, therefore, is to comprehensively parse and accurately model these expressions. To address this, we decompose Query Understanding into two core sub-tasks: Entity Recognition and Intent Recognition (entity-relation recognition). Our analysis begins with single-dimensional queries and progressively extends to more complex, multi-dimensional ones.

\textbf{(1) Entity recognition.} The objective of Entity Recognition is to extract all entities from a query that represent user needs, such as specific product attributes or item types. Based on their inherent difficulty and dependence on domain knowledge, these recognition tasks can be broadly categorized into the following scenarios:
\begin{itemize}
    \item \textbf{Knowledge-free scenarios.} the entities represented users need can be directly extracted from surface forms, with simple syntax and semantics, and almost no background knowledge is required.
    
    \item \textbf{General world knowledge scenarios.} While general semantic understanding is foundational, accurate interpretation in these cases hinges on substantial world knowledge. For instance, in ``pearl white'' or ``ivory whit'', the user's intent is a specific shade of white, not the literal objects ``pearl'' or ``ivory''. Similarly, ``phone gas station'' is a metaphorical expression for a ``power bank'', where ``gas station'' signifies a source of energy. Such examples demonstrate that higher-level semantic reasoning and world knowledge are tightly intertwined, rendering an approach based solely on pure semantic parsing insufficient.
    \item \textbf{Domain-specific knowledge scenarios.} These requirements involve niche characteristic attributes or derived attributes that cannot be inferred from surface forms alone and must rely on structured domain knowledge. Typical examples include domain-specific jargon in vertical e-commerce, shop/brand taxonomies, model coding rules, and category-specific descriptions of functions or effects (e.g., mouthfeel descriptions for tea or wine).
    \item \textbf{Other challenges: misspellings and aliases.} Users often express their needs using non-standard language, such as typos, homophones, aliases, or industry slang. Effectively addressing this requires a tighter integration of general and domain knowledge, demanding that the model possess robust capabilities for spelling correction, variant normalization, and alias alignment.
\end{itemize}

\begin{figure}[htbp]
    \centering
    \includegraphics[width=1\linewidth]{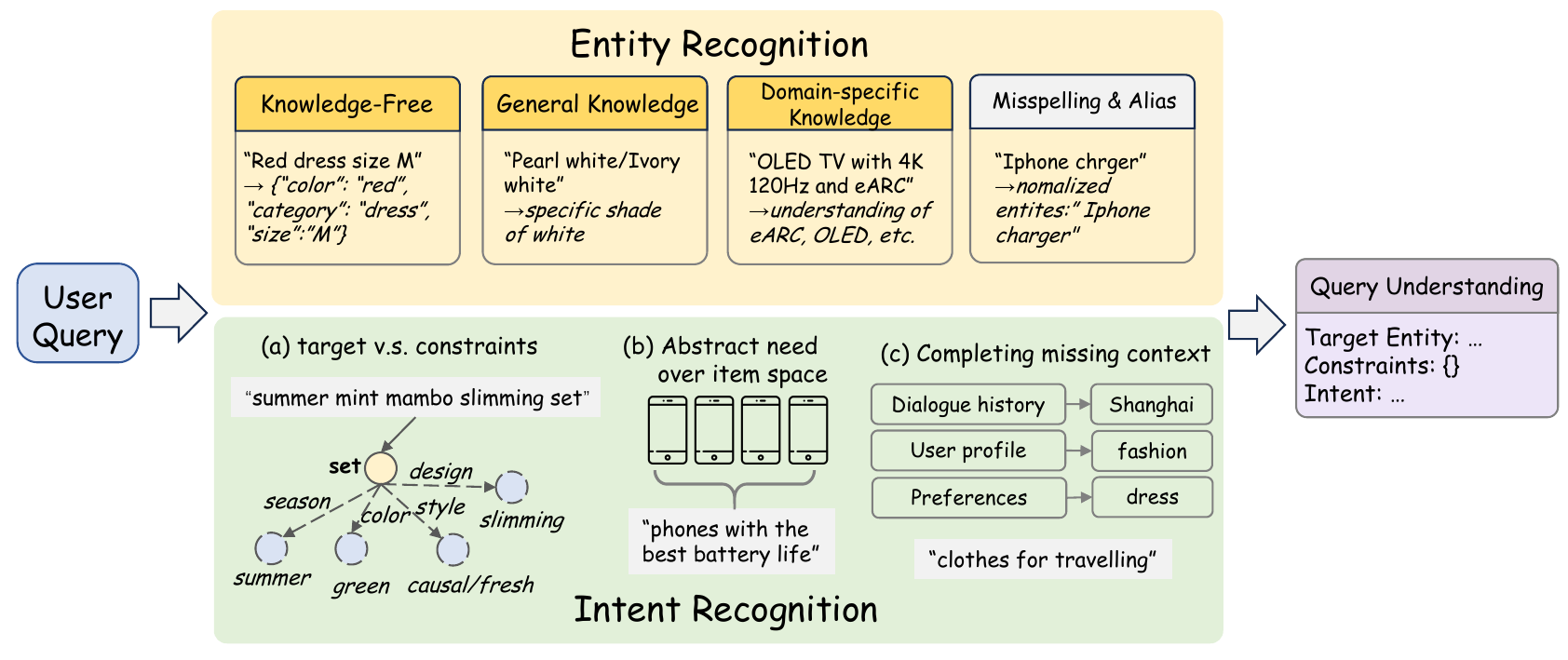}
    \caption{Comprehensive Overview of query understanding}
    \label{fig:placeholder}
\end{figure}

\textbf{(2) Intent recognition}
Intent recognition is the task of discerning whether a query is transactional (seeking an item) or informational (seeking related content such as reviews, comparisons, or usage guides). It also involves distinguishing the core subject of the query from its modifiers and constraints. This entire process is highly dependent on sophisticated reasoning. Key challenges in this area include:
\begin{itemize}
    \item Accurately identifying the purchase purpose and item usage, thereby locating the true target of the need and distinguishing whether the query is describing derived attributes or directly referring to specific items. For example, in ``summer mint mambo slimming set'', the target item is ``set", while the remaining modifiers serve as attribute constraints covering scenario (summer), function (slimming), brand (mambo), and style (mint).
    \item Handling abstract needs that do not point to specific items but to categories or domains. For instance, ``phones with the best battery life'' does not specify a concrete item, but poses a ranking or filtering request over the entire item space. The system must precisely capture the semantic meaning of the derived attribute ``battery life'' in the phone domain and its operational measurement, and then construct reasonable ranking or filtering rules to support inference and comparison over candidate sets.
    \item Mitigating ambiguity caused by missing context. In real interactions, users often do not specify all constraints or preferences within a single query. The system must therefore exploit dialogue history, user profiles, and default preferences to complete and disambiguate implicit context, obtaining a more fine-grained intent representation that supports reliable downstream relevance judgment.
\end{itemize}

\textbf{(3) From single-dimensional to multi-dimensional needs.}
In most e-commerce scenarios, user queries simultaneously encode multiple dimensions of needs (e.g., brand + price range + function + target user group). The system must therefore go beyond accurately identifying single-dimensional needs and, from a joint multi-dimensional perspective, fully extract and structure all need points. This provides the basis for subsequent path construction and relevance judgment. It is important to emphasize that the dependence on multi-modal capabilities should not be decided at the query-need level. Query-level tasks focus on entity recognition and intent recognition. The degree of multi-modal reliance should instead be determined by the properties of the identified attributes themselves, and discussed together with item attributes in the subsequent path design and judgment stages.

\subsubsection{Path Modeling from Query to Item}

In the overall framework, relevance judgment is abstracted as a process of constructing and executing paths from a query to an item. The system first maps the user query into structured retrieval and constraint paths, then performs candidate-item filtering and judgment along these paths. From this perspective, relevance judgment decomposes into two interrelated sub-tasks: (1) constructing correct and complete paths in the semantic and attribute space, which involves systematic modeling of query intent, constraints, and their mappings to item attributes; and (2) executing stable and reliable item judgment along these established paths to ensure consistent and reproducible assessment. The technical challenges primarily stem from the complexity and uncertainty inherent in these two processes. Below, we use the query \textbf{``Summer Mint Mambo Slimming Set''} to illustrate the modeling process.

\begin{figure}[htbp]
    \centering
    \includegraphics[width=1\linewidth]{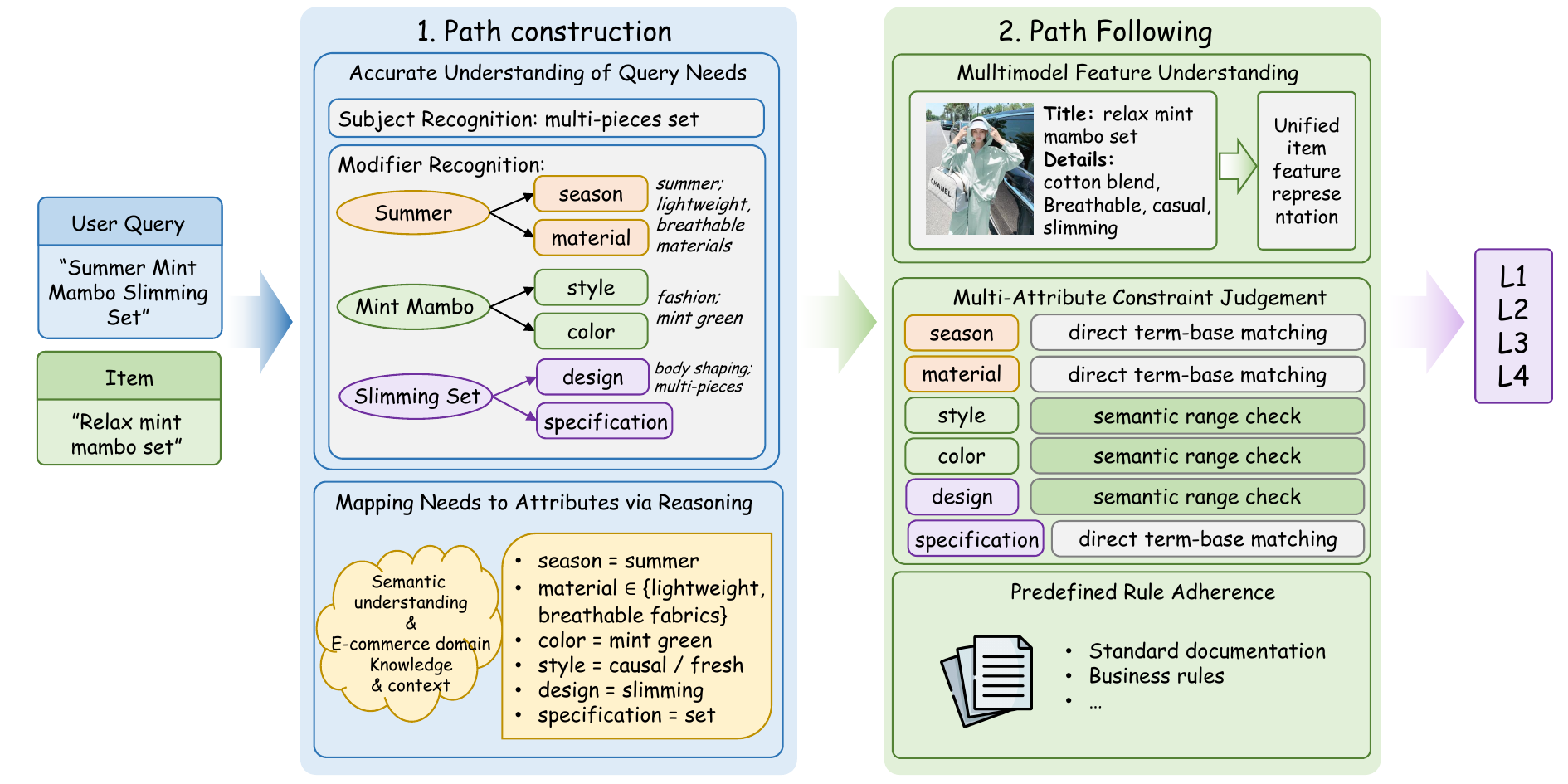}
    \caption{Overview of the Relevance Discrimination Process: (1) Path Construction, (2) Path Following}
    \label{fig:path}
\end{figure}

\paragraph{1. Path Construction} This phase refers to mapping query requirements to specific item attributes. The key challenge lies in accurately aligning the need points in the query with the specific attribute dimensions of the item. This process involves two main levels:

\textbf{(1) Accurate Understanding of Query Needs.} This level ensures the direction and precision of the path. \begin{itemize} \item \textbf{Subject Recognition:} This determines the target type (e.g., goods, content, or reviews).  In this example, the system identifies the target as an apparel set. \item \textbf{Modifier Recognition:} This parses modifiers as specific path constraints. Each identified need point acts as a constraint, and their intersection converges on the target. In this example: ``Summer'' constrains \textit{season} and \textit{material}; ``Mint Mambo'' constrains \textit{style} and \textit{color}; and ``Slimming Set'' constrains \textit{silhouette} and \textit{specifications}. \end{itemize}

\textbf{(2) Mapping Needs to Attributes via Reasoning.} While some constraints are explicit, many others—particularly those related to subjective attributes like material, style, and color—are often implicit and lack direct matching rules.
Therefore, the model must integrate semantic understanding and logical reasoning with e-commerce domain knowledge and contextual cues. This allows it to flexibly align natural language expressions with the structured attribute space.

\paragraph{2. Path Following} This phase refers to the execution of specific judgment logic by the model at the attribute-item level. Once query needs are mapped to path constraints, the relevance model performs feature extraction and verification along these paths. This process involves three key steps:

\begin{itemize} \item \textbf{Multimodal Feature Understanding:} The model must jointly utilize textual and visual information to extract item attributes to cover different types of constraint paths. For instance, for the ''Mint Mambo'' constraint, the model focuses on extracting visual features (color and style) for matching; whereas for the Set'' constraint'', it emphasizes extracting textual features related to specifications. \item \textbf{Multi-Attribute Constraint Judgment:} The model executes differentiated verification logic based on attribute types. For intrinsic attributes (e.g., specifications), the model performs direct rule-based or term-based matching. For derived attributes (e.g., ``Slimming''), the model checks whether the item's features fall within the valid semantic scope defined during the path construction phase. \item \textbf{Predefined Rule Adherence:} To ensure standardized, unified, and reproducible judgment results, the model strictly follows standard documentation and business rules. By unifying the attribute definition hierarchy and templating common query mapping relationships, the model converts judgment rules into executable configurations, guaranteeing reproducible assessment results across different items and scenarios. \end{itemize}
\begin{table}[!t]
    \centering
    \caption{Mapping Core Capabilities to Relevance Analysis Components}
    \label{tab:capabilities_mapping}
    \renewcommand{\arraystretch}{1.3}
    \begin{tabular}{p{0.25\textwidth} p{0.3\textwidth} p{0.38\textwidth}}
        \toprule
        \textbf{Core Capability} & \textbf{Analysis Component} & \textbf{Specific Application Context} \\
        \midrule
        \multirow{3}{=}{\textbf{Reasoning and Knowledge Integration}} 
        & \textbf{Query Understanding} & Resolving domain-specific entities (e.g., jargon, metaphors) and inferring user intent beyond surface text. \\
        & \textbf{Item Understanding} & Interpreting semantic meanings of \textit{derived attributes} (e.g., "usage scenarios," "feelings"). \\
        & \textbf{Path Construction} & Mapping abstract natural language needs (e.g., "for summer trips") to structured attribute combinations. \\
        \midrule
        \multirow{2}{=}{\textbf{Multi-modal Understanding}} 
        & \textbf{Item Understanding} & Extracting features for attributes that rely on presentation modality, such as \textit{style}, \textit{color}, and \textit{material}. \\
        & \textbf{Path Following} & Performing "Multimodal Feature Alignment" to verify visual constraints (e.g., matching a "Mint Mambo" visual style). \\
        \midrule
        \multirow{2}{=}{\textbf{Complex Rule Adherence}} 
        & \textbf{Item Understanding} & Handling \textit{inherent attributes} (e.g., specifications, brands) that require strict exact matching. \\
        & \textbf{Path Following} & Executing "Predefined Rule Adherence" to align model judgments with business SOPs and ensure reproducibility. \\
        \bottomrule
    \end{tabular}
\end{table}

\subsubsection{Core Capabilities for Relevance Judgment}

Based on the decomposition of item understanding, query understanding, and path modeling, we can synthesize the essential requirements for a robust relevance model. To effectively navigate the vast item corpus and satisfy complex user intents—ranging from direct specification to abstract reasoning—the model must possess a sophisticated, composite set of capabilities. We summarize these core requirements as:

\begin{itemize}
    \item \textbf{Reasoning and Knowledge Integration:} The ability to leverage external knowledge and logical inference to resolve ambiguity in query understanding and accurately map abstract intents during path construction.
    \item \textbf{Multi-modal Understanding and Matching:} The capacity to parse, align, and match attributes across textual and visual modalities, which is critical for handling image-heavy constraints in item understanding and path following.
    \item \textbf{Complex Rule Adherence:} The fidelity to learn and precisely execute the complex, often subtle, judgment rules required to ensure the standardization and reproducibility of the path following process.
\end{itemize}

Table~\ref{tab:capabilities_mapping} details how these core capabilities map to the specific components and challenges identified in our analysis.

\section{Method}
\subsection{Overview}

\textbf{Evolution of Large Model Training Paradigms for Relevance.} From a formulation perspective, the relevance task can be regarded as a classification task. Early works incorporated the task requirements as prompts and the annotated results as responses, performing SFT loss to fine-tune LLMs as follows, which serves as our iterative baseline.
\begin{equation}
\label{equ:SFT}
\mathcal{L}_{\text{SFT}} = -\sum_{i=1}^{N} \log P(y_i | x_i; \theta)
\end{equation}
However, simple SFT was soon proven to suffer from out-of-distribution performance degradation. Researchers discovered that leveraging the model's CoT capabilities—prompting the model to reason before outputting answers—could effectively enhance model robustness and performance on challenging samples. Due to constraints in computational resources and application scenarios, models with relatively modest parameter sizes are typically selected for training, which implies insufficient reasoning capabilities and instruction-following abilities under the prompting-only paradigm. Therefore, the current mainstream approach is to leverage powerful teacher models (such as GPT-5 and DeepSeek-R1\citep{deepseek}) to synthesize CoT data, and then distill the reasoning capabilities into relatively smaller models through SFT. Motivated by the remarkable achievements of reinforcement learning in LLM training, some works further incorporate an RL stage following CoT distillation, employing either offline RL methods such as Direct Preference Optimization(DPO)\citep{DPO} and KTO\citep{KTO}, or online RL methods such as Group Relative Policy Optimization(GRPO)\citep{GRPO}. 

\textbf{LORE's Training Paradigm.} Prior to formal training, we first conducted some necessary preliminary explorations, mainly covering three aspects: base model selection, prompt tuning, and feature infrastructure. These form the foundation of our training and will be elaborated in detail in Section~\ref{sec:3.2}. 

Our complete training paradigm consists of two stages: SFT and RL stage. In the SFT stage(Section~\ref{sec:3.3}), we construct a progressive CoT synthesis pipeline aligned with the two-stage discrimination framework. The first step performs knowledge injection and reasoning to accomplish path construction, followed by two subsequent steps that execute multi-modal attribute matching and rule-aware discrimination to complete path following. Furthermore, we also discuss issues pertaining to dataset construction, denoising, and training data proportions.
During this stage, we primarily focus on the model's pass@8 metric as shown in equation \ref{equ:pass@8}, which serves as an indicator for probing the model's capability boundaries. When performing eight inference iterations, we fix the model temperature at 1.0 to maintain a balance between accuracy and diversity. The model proceeds to RL only when pass@8 surpasses the baseline, indicating sufficient cold start, as RL amplifies correct answer probabilities but cannot compensate for consistently poor outputs. 
\begin{equation}
\label{equ:pass@8}
Pass@8 = 1 - \prod_{i=1}^{8} \left(1 - p_i\right)
\end{equation}
In the RL stage(Section~\ref{sec:3.4}), we design verifiable outcome rewards based on the characteristics of the relevance task, guiding the model to align all capabilities acquired in the previous step with human preferences and prune incorrect reasoning paths. Additionally, we conducted extensive explorations on sampling strategies, entropy optimization, and importance sampling granularity. During this stage, we focus on the model's pass@1 capability, as RL trades breadth for precision by converting pass@8 performance into first-attempt accuracy. To ensure result stability, we employ greedy decoding when computing the pass@1 metric.

Following the training phase, we develop a comprehensive and challenging evaluation benchmark designed to rigorously assess model capabilities(Section~\ref{sec:3.5}).
Given the powerful capabilities of LLMs and the diversity of e-commerce data, we designed our evaluation benchmark to ensure sample comprehensiveness while incorporating targeted hard sample mining based on the core capabilities identified in Section~\ref{prel}. 

\subsection{Preliminary Exploration}
\label{sec:3.2}
\begin{figure}[htbp]
  \centering
  \includegraphics[width=1\textwidth]{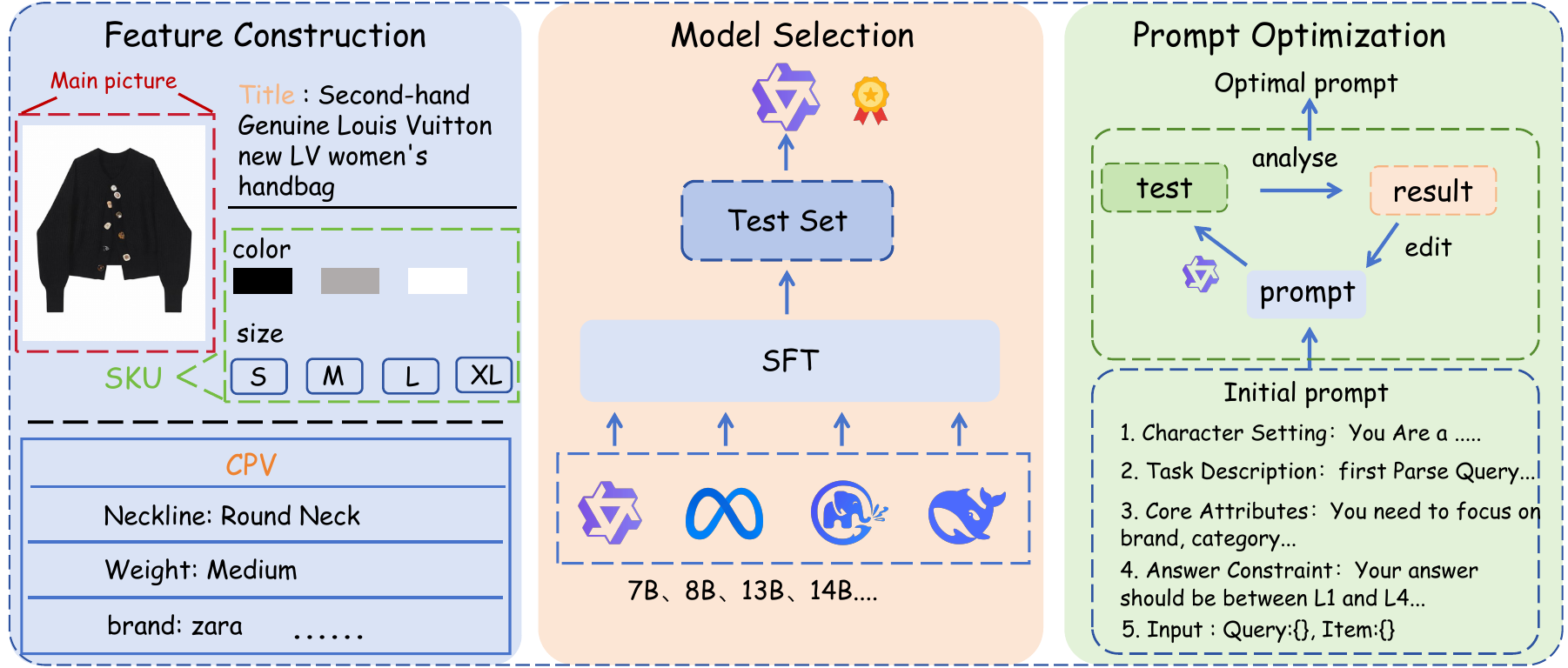}
  \caption{Preliminary exploration prior to training, encompassing: (1) Feature construction, (2) Model selection, and (3) Prompt optimization.}
  \label{fig:3.1}
\end{figure}

In this section, we conducted a series of necessary preliminary explorations, including three parts: feature construction, model selection, and prompt optimization as shown in Fig \ref{fig:3.1}.

\subsubsection{Feature Construction}
\label{sec:3.2.1}
Guided by the theoretical framework established in Section~\ref{prel}, effective relevance modeling necessitates a comprehensive construction of the item's attribute space. Thus, we select features to endow the model with the necessary perception and reasoning skills spanning both visual and textual modalities.

First, to address the need for visual attributes, such as stylistic details, cuts and patterns, we incorporate the item's Main Product Image. This visual input is essential for ``image-heavy'' queries where textual descriptions alone are insufficient.
For textual attributes, we begin by introducing Category-Property-Value (CPV) triples, which provides a rich, semi-structured summary of an item's core characteristics.
Considering that CPV information contains some redundant items that are irrelevant for relevance judgment, we introduce a pipeline combining prior knowledge with LLMs for filtering. Specifically, CPV information is presented in key-value pairs, and there exists key co-occurrence across similar products. For example, both "eye cream" and "face cream" products typically contain the key "brand". Therefore, we conducted statistical counting based on a large volume of product data to obtain medium-to-high frequency keys. Subsequently, we prompted a powerful LLM (Qwen3-235B-Instruct) with task definition, the considered attributes and each key, asking the model to judge whether each attribute is beneficial for the relevance task. To improve judgment accuracy, we performed eight sampling iterations on the model outputs and retained only those attributes where the self-consistency result was "true", which can be described as follows:
\begin{equation}
S_{CPV} = \left\{k\mid \sum_{i=1}^{8} (\text{Qwen}(k_i) = true) >= 4 \right\}
\end{equation}
Moreover, CPV information alone is insufficient, as it often fails to capture the fine-grained, purchasable variations—such as different colors, sizes, or package contents. To bridge this gap, we incorporate Stock Keeping Unit (SKU) information. SKU provide details for these specific variants, enabling our model to make precise assessments.

Even after the deduplication process, the resulting CPV data can still be noisy due to the inherent quality issues of the original item information. Meanwhile, SKU information often suffers from textual repetition and redundancy. To validate the effectiveness of the features we introduced, we designed a set of incremental analysis experiments. These experiments use the product title as a performance baseline and progressively add other features to evaluate their individual impact on model performance.
As shown in Table \ref{tab:tezheng1}, introducing both CPV and SKU features on top of the title improves model performance. This result indicates that even in the presence of partial redundancy and noise, input information can still yield significant performance gains as long as it contains sufficiently relevant and stable signals.
\begin{table}[h]
\centering
\caption{Experimental results of item feature enhancement.}
\renewcommand{\arraystretch}{1.2}  
\setlength{\tabcolsep}{1.2\tabcolsep}
\begin{tabular}{lccc}
\toprule
Models & pass@1  \\
\midrule
base(Title)& 0.847 \\
\hline
base + CPV &  0.855\\
\hline
base + CPV + SKU & 0.871  \\
\bottomrule
\end{tabular}
\label{tab:tezheng1}
\end{table}


\subsubsection{Model Selection}
\label{sec:model_selection}
To select an optimal base model, we balance two critical factors: the capability to execute the complex reasoning and knowledge integration detailed in Section~\ref{prel}, and the efficiency required for practical deployment.

We selected a series of excellent open-source models with parameters ranging from 7B to 14B. This size range ensures sufficient capability while offering the advantage of single-GPU training and inference. Subsequently, we selected a portion from the existing annotated data, dividing it into training and validation sets. Using the basic relevance task definition as the prompt, we performed simple fine-tuning as mentioned in Section 3.1 for each model on the training set and validated the performance. Finally, the Qwen2.5-7B model achieved superior performance among all evaluated models while maintaining high inference efficiency. Consequently, we selected it as the base model for our framework.


\subsubsection{Prompt Optimization}
\label{sec:3.2.3}
In domain-specific SFT, prompts are typically constructed using a fixed template containing task definitions and requirements, populated with query and item data from individual samples. To fully activate the base model's core capabilities, it is crucial to embed key elements such as relevance definitions, judgment rules, and output formats. Furthermore, to align the model with our proposed two-stage discrimination framework, we integrated structured instructions requiring the model to first analyze the query's intent and attribute requirements, and subsequently extract relevant item attributes to render a judgment based on established rules.

While detailed instructions generally enhance performance in reasoning tasks, this correlation does not strictly apply to SFT scenarios, where models acquire capabilities primarily through learning from response data rather than relying solely on prompt instructions. To identify the optimal configuration, we evaluated three prompt designs: a long version (about 7,000 tokens) containing exhaustive rules; a mid version (about 800 tokens) retaining only essential instructions; and a short version consisting solely of a basic role definition (e.g., ``You are an e-commerce relevance assessment expert...'').

\begin{table}[h]
\centering
\vspace{-0.4cm}
\caption{Experimental results of SFT across different prompts.}
\renewcommand{\arraystretch}{1.2}  
\setlength{\tabcolsep}{1.2\tabcolsep}
\begin{tabular}{lc}
\toprule
Prompt  & pass@1  \\
\midrule
short prompt & 0.861 \\
\hline
mid prompt&  0.871\\
\hline
long prompt& 0.868  \\
\bottomrule
\end{tabular}
\label{tab:prompt}
\end{table}

We validated using the training and test sets mentioned in Section~\ref{sec:model_selection}, with results shown in Table \ref{tab:prompt}. 
Model performance showed minimal variance across the three prompt versions (within 1\%), indicating that in SFT scenarios, model improvements stem primarily from learning ground truth responses rather than relying on prompt engineering. Specifically, the short prompt yielded the poorest results as expected; its lack of essential task information likely forced the model to merely fit the data distribution without true understanding. Surprisingly, the long prompt, despite its rich detail, underperformed the streamlined mid prompt. We hypothesize that excessive length causes attention dispersion, thereby complicating the discrimination process. The results demonstrate that a concise prompt containing only core information yields optimal performance. Consequently, we adopted the mid-length prompt configuration for all subsequent fine-tuning experiments.

\subsection{SFT: Comprehensive Reasoning Capability Injection}
\label{sec:3.3}
\subsubsection{Data construction}
This phase constitutes the model cold start stage, designed to establish the model's fundamental discriminative capabilities. Achieving this requires a training dataset characterized by two core attributes: comprehensiveness and low noise. To ensure comprehensiveness, we employ a scientific sampling strategy that adequately covers diverse e-commerce data distributions, thereby mitigating potential oversampling or undersampling bias. Addressing noise is equally critical; since manual annotation inevitably introduces errors—with empirical tests showing an accuracy of only 95\%—we implemented a robust data cleaning pipeline to systematically reduce noise and enhance overall dataset quality.
\begin{figure}[htbp]
  \centering
  \includegraphics[width=1\textwidth]{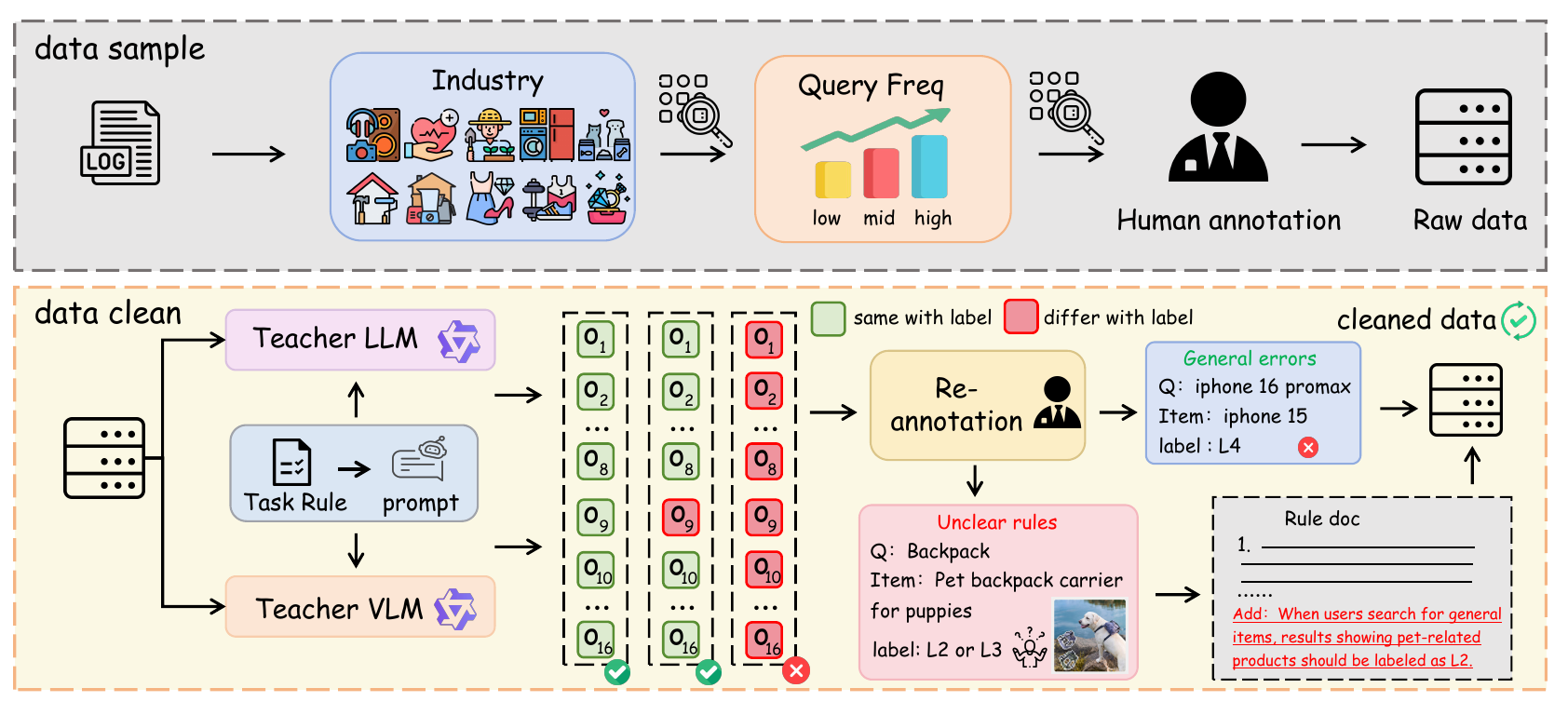}
  \caption{Framework for SFT training data sampling and cleaning.}
  \label{fig:SFTdata}
\end{figure}

\textbf{Data sampling.} Based on Taobao search logs, we first categorize the data by industry-level such as consumer electronics, apparel, etc., and conduct stratified sampling according to the proportion of each industry. The purpose of this step is to ensure that the training data cover attributes and features across all industries. Next, we divide the data for each industry into high/medium/low frequency categories based on query and conduct sampling according to their respective proportions, ensuring that the dataset encompasses both common high-frequency queries and long-tail difficult samples. Subsequently, the data is assigned to specialized annotators for labeling.

\textbf{Data cleaning.} 
Errors in manual annotation primarily stem from two sources. First, human oversight often leads to the omission of details, a tendency particularly pronounced when processing long, multi-attribute queries. Second, rule ambiguity poses a challenge; as manifestations of subjective user preferences, existing rules struggle to exhaustively cover all edge cases. To address these issues, we leverage powerful language models for data filtering: (1) The attribute matching model executes strict matching strategies, demonstrating superior fine-grained feature capture compared to manual annotation; (2) By injecting comprehensive rules into prompts to guide reasoning, the model could effectively identify and filter out corner cases that fall beyond the scope of standard rules.

Specifically, we prompt a powerful text model, Qwen3-235B-Instruct, with detailed definitions of relevance tasks and rules, instructing the model to make 8 separate judgments on each pair $<Q_i, I_i>$. Subsequently, we prompt a powerful VLM, Qwen2.5-72B-VL with the same settings and additionally incorporated product main images to obtain eight judgment results for each $<Q_i, I_i>$ pair. We merged the results from both inference rounds, totaling 16 judgments, and filtered out samples where none of the judgments agreed with the manual annotations, as these are highly likely to be incorrectly labeled samples or Extremely difficult samples. This process can be represented as follows: 
\begin{equation}
S_{noise} = \left\{ (Q_i, I_i) \mid \sum_{j=1}^{16} \mathbb{1}[J_j(Q_i, I_i) = y_i] = 0 \right\}
\end{equation}

The rationale for employing a dual-model approach is to synergize the text model's robust long-context comprehension with the VLM's multimodal capabilities, thereby minimizing both false positives and false negatives during filtering. Subsequently, suspicious samples identified by this process undergo manual review and are categorized for remediation: clear annotation errors are corrected directly, while discrepancies arising from rule ambiguity trigger the formulation of new guidelines based on case analysis, facilitating the iterative refinement of the annotation rule system. Table \ref{tab:data} presents the post-cleaning results. The process resulted in label modifications for 6.1\% of the samples, elevating overall data quality from 95\% to 99\%.
\begin{table}[h]
\centering
\caption{Comparison of training set quality before and after denoising.}
\begin{tabular}{lcc}
\toprule
data & correction rate & Accuracy\\
\midrule 
before cleaning & - &  95\% \\
after cleaning & 6.1\% & 99\%   \\
\bottomrule
\end{tabular}
\label{tab:data}
\end{table}

\subsubsection{Multi-dimensional Chain-of-Thought Synthesis}
\begin{figure}[htbp]
  \centering
  \includegraphics[width=1\textwidth]{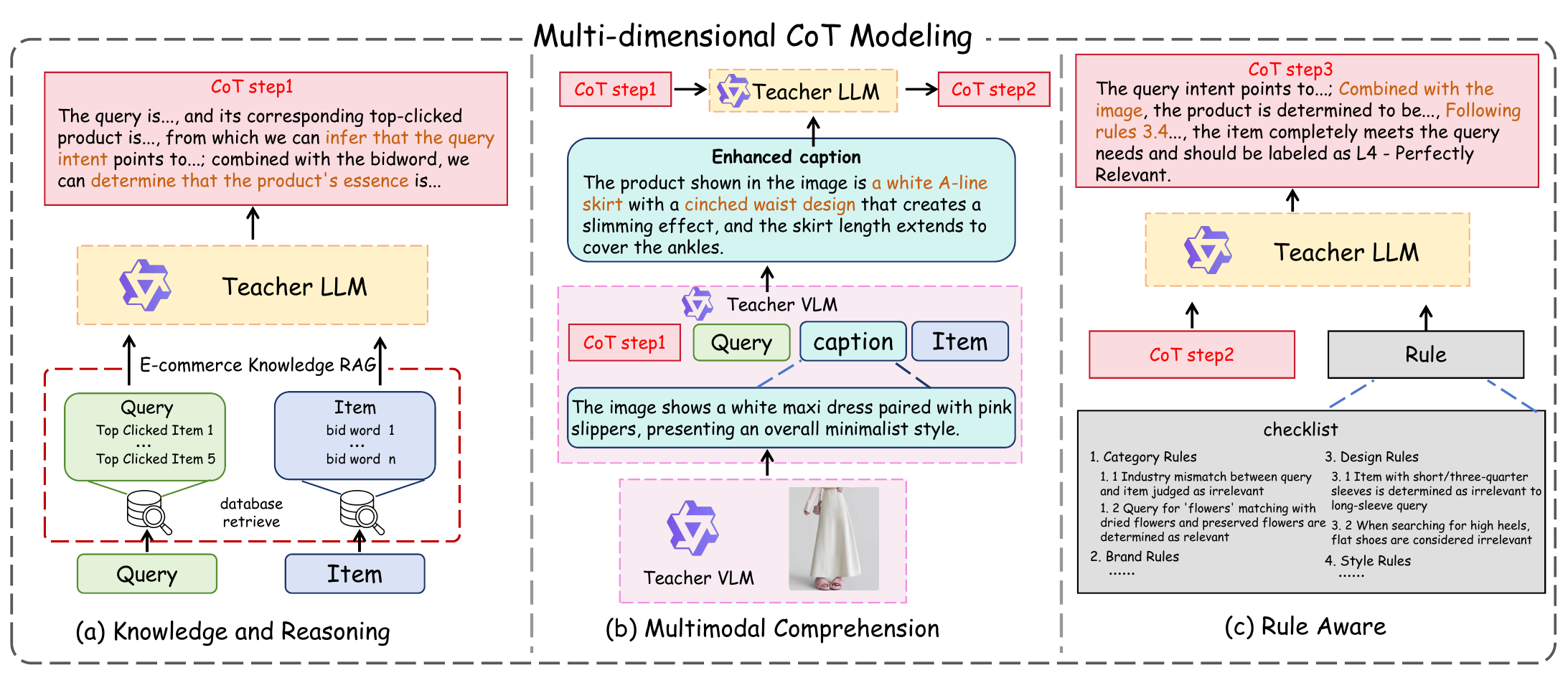}
  \caption{Progressive multi-dimensional CoT synthesis framework.}
  \label{fig:3.2}
\end{figure}

To enable the model to possess comprehensive and in-depth reasoning capabilities, we proposes a progressive CoT synthesis pipeline as shown in Fig~\ref{fig:3.2}. This pipeline adopts a stage-wise construction strategy, sequentially encompassing three progressive stages: (1)\textbf{Knowledge Injection \& Reasoning, (2)Multimodal Comprehension, and (3)Rule Aware}. Knowledge Injection \& Reasoning completes the Path Construction stage, which parses query intent into attributes comparable with items. Multimodal Comprehension and Rule Awareness then complete the Path Following stage, which evaluates the degree to which items satisfy the query based on item information and discrimination rules.

\textbf{Step 1: Knowledge Injection \& Reasoning.} knowledge serves as the cognitive foundation for interpreting user queries and item attributes, necessitating its injection at the initial stage. By leveraging this knowledge base, the model can achieve a profound understanding of query intent, accurately infer core user requirements, and extract relevant attribute features from item information. As outlined in Section 2.3, we categorize knowledge into general world knowledge and e-commerce knowledge. Regarding the former, existing LLMs have acquired substantial general knowledge through large-scale pre-training, enabling them to grasp common concepts and entity relationships. However, e-commerce scenarios frequently involve specialized terminology and phrasing whose meanings cannot be fully comprehended based solely on literal semantics. 

Such challenges can be primarily categorized into the following two dimensions:
(1) Long-tail Queries: These queries explicitly refer to highly niche items. However, existing models often lack the necessary prior knowledge, making it difficult to establish effective semantic mappings. For instance, consider the query "strange things are also loved by someone." While this refers to a specific book title, a model relying solely on literal semantics might incorrectly associate it with "novel and interesting toys." Without specific knowledge—such as brand slang, product nicknames, or marketing terminology—it is difficult for the model to make accurate relevance judgments.
(2) Semantically Ambiguous Items: Certain items fall outside the model's knowledge scope due to their extremely niche nature. For example, a term like "lying gate" cannot be accurately identified or categorized based on textual title information alone; it requires a deeper contextual understanding of product attributes.
Ultimately, both cases necessitate supplementary domain-specific e-commerce knowledge for accurate interpretation.

To address these issues, we introduced a Retrieval-Augmented Generation (RAG) mechanism and constructed a dynamic knowledge database specifically designed for e-commerce scenarios: (1) Query-side Knowledge Enhancement: To inject the model with prior knowledge of items targeted by queries, we introduce the titles of the top five items with the highest historical click-through rates for each query as contextual information. These high-click items can effectively reflect users' actual search intent and item preference characteristics. (2) Item-side Knowledge Enhancement: To inject the model with more comprehensive item information, We incorporate merchant-provided "selling points", which can be regarded as merchants' refined descriptions of core product features and highlights, containing key attributes, applicable scenarios, and differentiated advantages of the products. We retrieved augmented knowledge for each data sample and concatenated it with query $Q_i$ and Item $I_i$ to construct the context as follows, where $K_Q$ and $K_I$ represent the e-commerce augmented information.
\begin{equation}
context_i = concat(Q_i,K_Q,I_i,K_I)
\end{equation}

Upon completion of knowledge augmentation, the model has acquired the fundamental capability to perform deep-level reasoning on query intent and attributes. We employs the Qwen3-235B-Instruct model to generate reasoning paths. Through carefully designed prompt templates, the model is guided to execute multi-level reasoning analysis: first conducts semantic understanding of the query by integrating its inherent knowledge with e-commerce domain knowledge, inferring users' true intent and extracting specific attribute features; subsequently, it performs systematic analysis of item information, deconstructing the item's concrete attributes and establishing comparative mappings with the query. This process generates reasoning chains $CoT_{step1}$ that comprehensively understand both queries and items as follows:
\begin{equation}
\text{CoT}_{\text{step1}} = \arg\max_{y} P_{\theta}(y \mid concat(prompt_{LLM},context_i)
\end{equation}

\textbf{Step 2: Multimodal Comprehension.} 
The visual presentation of items often contains critical features that are difficult to fully express through text (such as style details, material texture). In light of this, this stage further introduces item image information, synthesizing attribute extraction chains of thought that can integrate both textual and visual modalities. Based on the following two considerations, we did not directly employ VLM for multimodal modeling: (1) Limited reasoning capability: Existing VLMs generally exhibit weaker reasoning capabilities compared to LLMs of equivalent parameter scale, a phenomenon that has been validated in multiple benchmark tests and our experiments. (2) Modality dependence bias: Direct fine-tuning of VLMs tends to cause the model to exhibit a "text shortcut" phenomenon, i.e., over-reliance on textual information while neglecting image features. The underlying cause lies in the inherent information density disparity between text and image modalities.

Therefore, we designed a caption-mediated two-stage modeling framework: in the first stage, we extract image semantics through VLM and generate textual descriptions; in the second stage, these descriptions are incorporated into the prompt as additional context for fine-tuning. However, conventional caption generation methods face inherent limitations. Given the complexity of visual content and the absence of task-specific guidance, VLMs struggle to effectively identify key information relevant to reasoning. This leads to the omission of critical visual cues—a challenge that cannot be fully addressed through simple prompt engineering. Consequently, we propose a multimodal CoT generation framework guided by relevance judgment.

Firstly, We prompt a Qwen2.5VL-72B model to generate a basic caption $C_{naive}$ for the image as follows:
\begin{equation}
C_{naive} = \arg\max_{y} P_{\theta}(y \mid prompt_{VL},I_i)
\end{equation}
Secondly, We integrate the original input $C_{naive}$, the $CoT_{step1}$ obtained from the previous section, and other contextual information into a structured prompt, and input it to Qwen2.5-VL-72B to regenerate the caption. This design enables the model to extract task-relevant visual information from the image in a targeted manner under the guidance of explicit reasoning key points, thereby generating more precise and task-oriented descriptions. 
\begin{equation}
C_{enhanced} = \arg\max_{y} P_{\theta}(y \mid Concat(prompt_{VL},Context_i,CoT_{step1},C_{naive}))
\end{equation}
Finally, The $C_{enhanced}$, along with $CoT_{step1}$ and contextual information, are injected into the prompt for Qwen2.5-235B-Instruct to produce the complete $CoT_{step2}$, which encompassing the complete relevance discrimination process.
\begin{equation}
\text{CoT}_{\text{step2}} = \arg\max_{y} P_{\theta}(y \mid Concat( prompt_{LLM},Context_i,CoT_{step1},C_{enhanced}))
\end{equation}

\textbf{Step3: Rule-Aware.} 
While the previous $CoT_{step2}$
effectively bridges the semantic gap, it lacks the objective verification mandated by Section 2.3. To address this, we synthesize rule-aware reasoning chains to enforce rigorous attribute matching. Rather than explicitly listing comprehensive rules—which would overwhelm the context window and hinder instruction following, we embed these constraints directly into the reasoning steps. This allows the model to internalize the judgment criteria implicitly through demonstration.
Specifically, we partition the rule system by industry to obtain industry-specific rule sets R. For each training sample, we construct a prompt comprising: (1) the rule set R, (2) the $context_i$ , (3) the ground-truth $L_i$, and (4) the previously generated $CoT_{step2}$. This prompt guides the Qwen3-235B-Instruct model to reverse-engineer a rule-based $CoT_{step2}$ for relevance judgment by integrating the complete rule framework, human annotations, and deep semantic comprehension of both query and item contexts.
\begin{equation}
\text{CoT}_{\text{step3}} = \arg\max_{y} P_{\theta}(y \mid Concat(prompt_{LLM}, context_i,CoT_{step2},L_i,R))
\end{equation}

\subsubsection{Distillation by SFT}
\label{sec:3.3.3}
To distill the multi-dimensional reasoning capabilities into a more compact model, we perform SFT on the base model using the synthesized $CoT_{step3}$ as training data, as formulated in Equation \ref{equ:SFT}. To facilitate result extraction and subsequent RLVR optimization, we organize the response data into the following format with separated reasoning process and results.
\begin{equation}
response = <think>reasoning\ process</think><answer> (L1|L2|L3|L4)</answer>
\end{equation}

Besides, a critical question remains to be addressed: as training progresses, model performance gradually approaches saturation, and further training risks overfitting. Therefore, determining the optimal SFT data scale is crucial, requiring the model to sufficiently learn reasoning patterns and acquire adequate e-commerce domain knowledge while avoiding generalization performance degradation caused by overfitting.

To address the aforementioned issue, we designed a data scale sensitivity analysis experiment: the training data was uniformly sampled with an initial proportion set at 10\%, followed by incremental increases in SFT data proportion at 10\% intervals (i.e., 10\%, 20\%, 30\%, ...). Throughout this process, we focused on the trends of the following two key metrics: (1) Format Accuracy: this metric reflects the model's mastery of reasoning patterns; (2) Pass@8 metric: this metric characterizes the theoretical upper bound of model performance. 

\begin{figure}[htbp]
  \centering
 \includegraphics[width=0.95\textwidth]{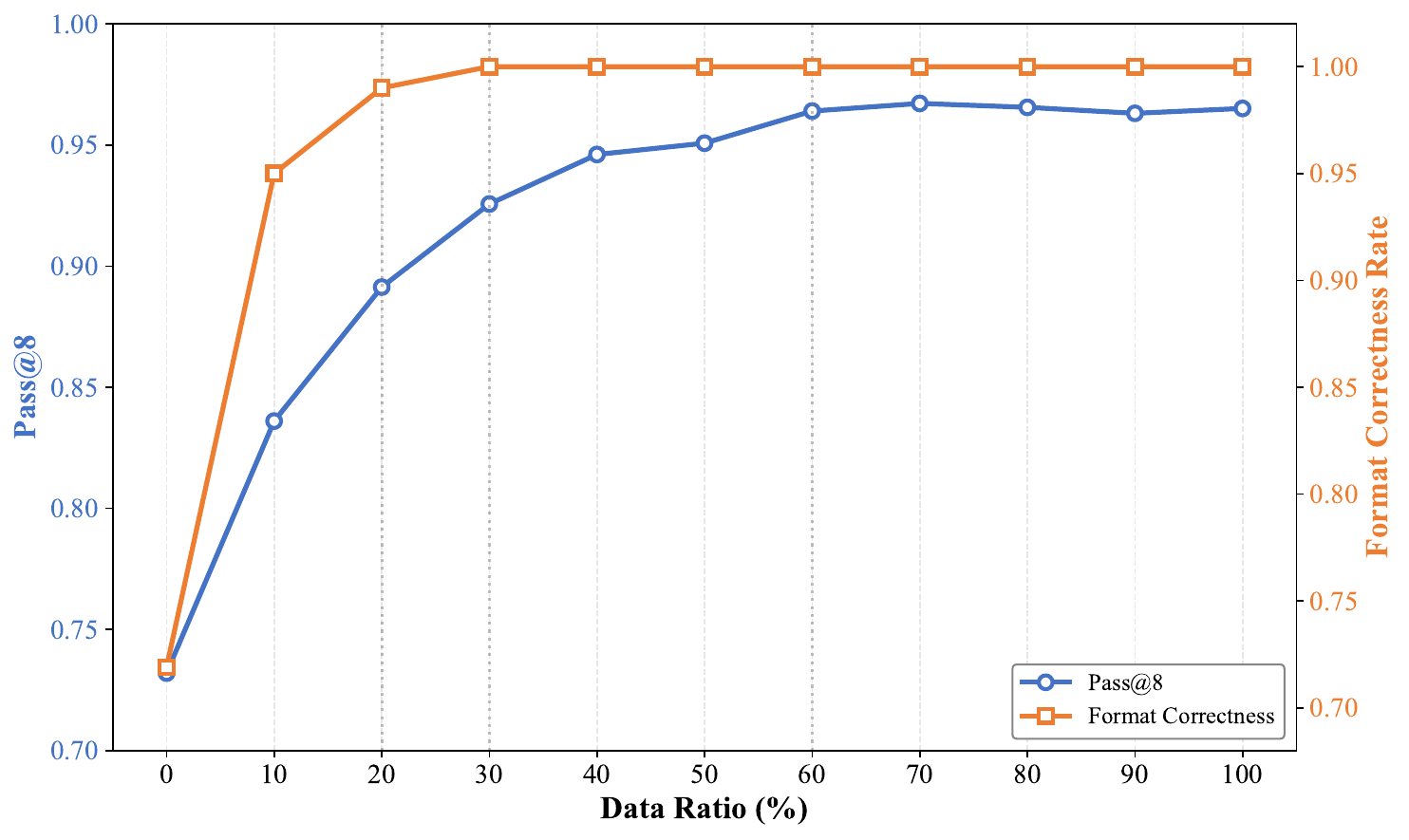}
  \caption{Evolution of pass@8 and format correctness rate with varying data proportions in SFT process.}
  \label{fig:SFTdata}
\end{figure}

Fig~\ref{fig:SFTdata} illustrates the trends in model metrics as the data proportion increases. The results demonstrate that the model can rapidly acquire the correct reasoning format in the early training stage, achieving a format accuracy exceeding 98\% with only 20\% of the training data. The Pass@8 metric exhibits rapid growth in the early training phase but subsequently displays a phenomenon of diminishing marginal returns. When approximately 60\% of the training data is utilized, the growth rate of this metric plateaus, showing no significant improvement. Based on these experimental observations, this study selects the 60\% data proportion as the optimal training scale, with which the cold-start model is trained.

\begin{table}[h]
\centering
\caption{Comparison of pass@8 between cold-start model with synthetic CoT distillation and vanilla SFT.}
\renewcommand{\arraystretch}{1.2}  
\setlength{\tabcolsep}{1.2\tabcolsep}
\begin{tabular}{cc}
\hline
Model  & pass@8 \\
\hline
vanilla SFT  & 0.937 \\
cold-start & 0.964 \\
\hline
\end{tabular}
\label{tab:model_performance}
\end{table}

Table~\ref{tab:model_performance} presents a comparison of the Pass@8 results between the cold-start model and the base model. The cold start model significantly outperforms the base model, indicating that multi-dimensional CoT modeling effectively elevates the model's performance ceiling, which provides ample room for subsequent RL optimization.

\subsection{RL: Relevance-Oriented Human Preference Alignment}
\label{sec:3.4}
In this section, we introduce reinforcement learning to guide the model to align with human preferences. By pruning erroneous reasoning paths, we effectively transform the reasoning capability demonstrated by the model across multiple samples (pass@8) into stable single performance (pass@1).

\subsubsection{Data construction}
\label{sec:3.4.1}
\begin{figure}[h]
\centering
\subfigure[before sample]{
\includegraphics[width=0.45\textwidth]{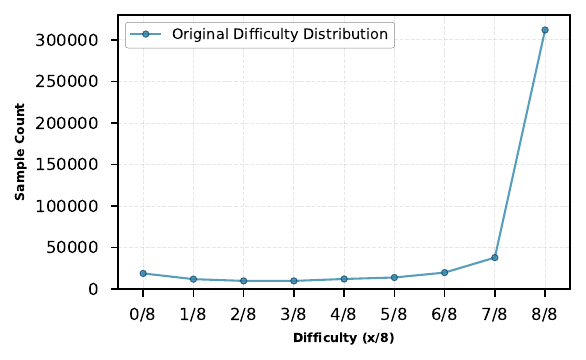}
}
\subfigure[after sample]{
    \includegraphics[width=0.45\textwidth]{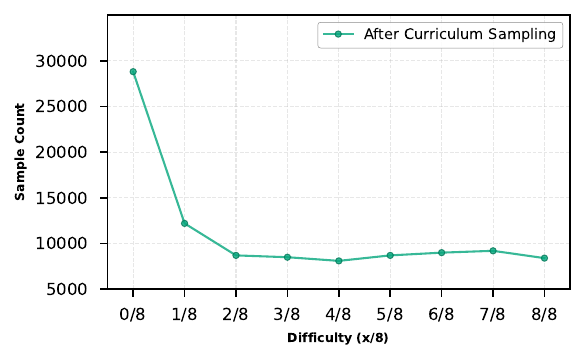}
}
\caption{Comparison of sample difficulty distribution before and after sampling.}
\label{fig:J}
\end{figure}

Given that the cold-start model has already mastered fundamental reasoning capabilities, subsequent reinforcement learning optimization urgently requires more challenging samples to achieve significant performance gains. Therefore, the core objective of data construction in the reinforcement learning phase lies in hard sample mining.
We use the cold-start model to perform 8-round sampling inference on the whole training data. As shown in Fig~\ref{fig:J} (a), the distribution of correct answer counts per sample exhibits a J-shaped distribution: there exists a large number of "easy samples" that the model has fully mastered (all 8 attempts correct). 
These samples have an advantage of 0, failing to provide effective gradient signals for RL training. To improve the data efficiency of RL training, we designed a difficulty-based sample filtering strategy as follows, where k denotes the number of correct answers and $\alpha$ denotes the downsampling rate.

\begin{equation}
\text{Sample Weight} = \begin{cases}1.0, & \text{if } k < 5  \\\alpha \in (0, 1), & \text{if } 5 \leq k < 8  \\0, & \text{if } k = 8 \end{cases}
\end{equation}
This strategy fully retains confusing samples (\textless5 correct), ensuring sufficient learning signals; completely removes saturated samples (8 correct), eliminating ineffective samples; proportionally downsamples relatively confident samples (5-7 correct), balancing the data distribution, ultimately forming an inverted J-shaped distribution as shown in Fig~\ref{fig:J}(b).

\subsubsection{Basic RL Framework}
\begin{figure}[htbp]
  \centering
 \includegraphics[width=0.95\textwidth]{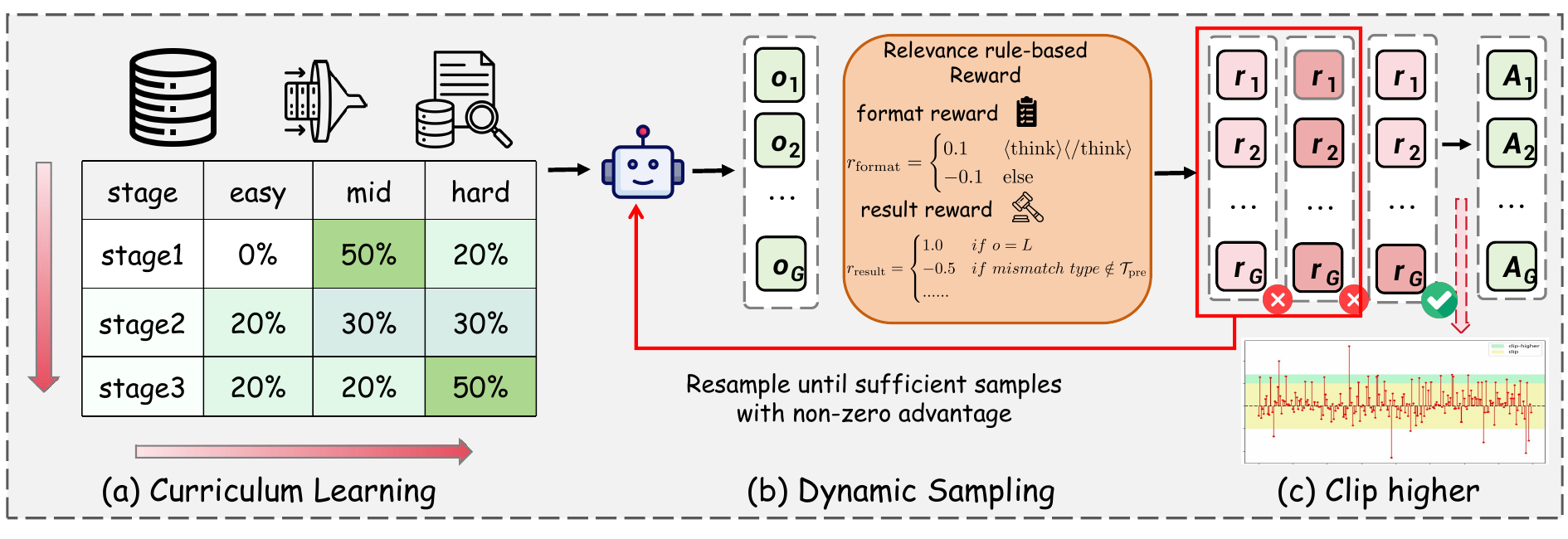}
  \caption{Overview of RLVR: basic framework and optimization strategies.}
  \label{fig:RL_basic}
\end{figure}

\textbf{KL-free GRPO Algorithm.} Fig~\ref{fig:RL_basic} illustrates our basic RL Framework. We adopt the GRPO algorithm to optimize the objective function as shown in Equation \ref{equ:grpo_loss}, which maximizes the advantage-weighted log probability expectation. The importance sampling ratio and advantage function are defined in Equation \ref{equ:ratio}. we removed the KL divergence regularization term in our actual implementation. This decision is based on the following considerations: after sufficient SFT warm-start, the model has already solidified its reasoning paradigm and structured output format. At this point, KL constraints would instead excessively limit the policy exploration space, hindering the model from discovering better reasoning paths. Therefore, we chose to allow the model to perform freer policy optimization based on its existing capabilities.

\begin{equation}
\label{equ:grpo_loss}
\mathcal{J}_{\text{GRPO}}(\theta) = \mathbb{E}_{x \sim \mathcal{D}, \{y_i\}_{i=1}^G \sim \pi_{\theta_{\text{old}}}(\cdot|x)} \left[ \frac{1}{G} \sum_{i=1}^{G} \frac{1}{|y_i|} \sum_{t=1}^{|y_i|} \min\left(w_{i,t}(\theta)\widehat{A}_{i,t}, \text{clip}(w_{i,t}(\theta), 1-\varepsilon, 1+\varepsilon)\widehat{A}_{i,t}\right) \right]
\end{equation}

\begin{equation}
w_{i,t} = \frac{\pi_\theta(y_{i,t}|x, y_{i,<t})}{\pi_{\theta_{\text{old}}}(y_{i,t}|x, y_{i,<t})}, \quad \widehat{A}_{i,t} = \frac{r_i - \text{mean}(\{r_j\}_{j=1}^G)}{\text{std}(\{r_j\}_{j=1}^G)}\label{equ:ratio}
\end{equation}

\textbf{Verifiable Outcome Reward Based on Relevance Labels.} We designed the reward system as follows by incorporating task-specific characteristics, which consists of two components: format reward and outcome reward. Where o represents the model output, L represents the ground truth given in L1-L4 levels, BinaryClass denotes the "relevant/irrelevant" outcome corresponding to the 4 levels, and $\mathcal{T}_{\text{pre}}$ represents the 18 types of attribute deficiency mismatches mentioned in Section \ref{prel}.

\begin{equation}
r = r_{\text{format}} + r_{\text{result}}
\end{equation}

\begin{equation}
r_{\text{format}} = \begin{cases} 
0.1 & \text{o follows }\texttt{<think>
</think>
<answer></answer>}\\[6pt]
-0.1 & else \\[6pt]
\end{cases}
\end{equation}

\begin{equation}
r_{\text{result}} = \begin{cases} 1.0 & \text{if } o = L \\[6pt]-0.5 & \text{if } o = L \text{ and mismatch type} \notin \mathcal{T}_{\text{pre}} \\[6pt]0.3 & \text{if } o \neq L \text{ and } \text{BinaryClass}(o) = \text{BinaryClass}(L) \\[6pt]-1.0 & \text{if } o \neq L \text{ and } \text{BinaryClass}(o) \neq \text{BinaryClass}(L) \\[6pt]-1.0 & \text{if } o \text{ is unparsable} \\[6pt]\end{cases}
\end{equation}
Through the reward mechanism design, we achieve three training objectives: (1) Format standardization: using special tokens to explicitly separate the reasoning process from the final answer, facilitating subsequent parsing and application; (2) Output space constraint: strictly limiting mismatch types (such as "price mismatch," "functionality mismatch," etc.) to within a predefined set, preventing the model from generating uncontrollable arbitrary outputs; (3) Progressive reward: when the model correctly identifies relevance but errs on fine-grained levels, still providing moderate positive rewards to mitigate the sparse reward problem in early training stages, accelerating convergence and stabilizing training.

\subsubsection{Training Strategy Optimization}
\label{sec:3.4.3}
\textbf{Curriculum Learning.\citep{cl}} We introduce a curriculum learning strategy in RL training, guiding the model to progressively improve its capabilities from simple to complex by gradually adjusting the sample difficulty distribution. Based on the 8-round inference statistics in Section~\ref{sec:3.4.1}, we categorize samples into three tiers according to the number of correct answers k as follows:
\begin{equation}
\text{Difficulty}(x) = \begin{cases}
\text{Easy} & \text{if } k(x) \in \{6, 7\} \\
\text{Medium} & \text{if } k(x) \in \{3, 4, 5\} \\
\text{Hard} & \text{if } k(x) \in \{0, 1, 2\}
\end{cases}
\end{equation}
Subsequently, we designed a three-stage curriculum learning scheme that guides model capability improvement by progressively increasing training data difficulty. The sample allocation strategy for each stage is detailed in Table \ref{tab:CL}. 

\begin{table}[h]
\centering
\caption{Progressive curriculum learning strategy with gradually increasing sample difficulty.}
\renewcommand{\arraystretch}{1.2}  
\setlength{\tabcolsep}{1.2\tabcolsep}
\begin{tabular}{cccc}
\toprule
stage  & easy & mid & hard \\
\midrule
stage 1 & 0\% & 50\% & 20\% \\
\hline
stage 2 & 20\% & 30\% & 30\% \\
\hline
stage 3 & 20\% & 20\% & 50\% \\
\bottomrule
\end{tabular}
\label{tab:CL}
\end{table}

The core philosophy behind this strategy encompasses three dimensions: (1) Rapid Bootstrap Principle: Stage 1 focuses primarily on medium-difficulty samples (rather than the easiest samples), enabling the model to quickly obtain effective positive feedback while maintaining a certain level of challenge, avoiding exploration difficulties caused by sparse rewards in early training. (2) Progressive Challenge Principle: Stages 2 and 3 gradually increase the proportion of difficult samples, guiding the model to transition from "solving common problems" to "tackling edge cases," achieving continuous expansion of capability boundaries. (3) Knowledge Retention Principle: The latter two stages continuously incorporate easy samples as "memory anchors" to prevent the model from forgetting old skills when learning new knowledge, while these high-confidence samples also help stabilize gradient variance during the training process. We conducted a systematic exploration of sample allocation for each stage through grid search, ultimately determining the proportion configuration shown in Table \ref{tab:CL} as the optimal solution.

\begin{figure}[htbp]
  \centering
 \includegraphics[width=1\textwidth]{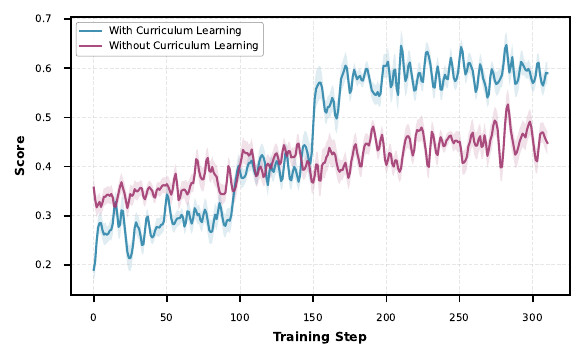}
  \caption{Comparison of reward curves between curriculum learning and non-curriculum learning.}
  \label{fig:课程学习}
\end{figure}

As illustrated in Fig~\ref{fig:课程学习}, the experimental group utilizing the curriculum learning strategy demonstrates faster convergence speed and more stable optimization trajectory in terms of both reward values and model performance improvement.

\textbf{Dynamic Sampling.\citep{DAPO}}
As RL training progresses, sample effectiveness gradually decays: after the model policy converges, an increasing number of samples exhibit consistent outcomes across multiple sampling attempts (either all correct or all incorrect), resulting in an advantage function $A_i=0$. Although these saturated samples participate in training, they fail to generate effective gradients, leading to computational resource waste and slowing the convergence speed. To address this issue, we introduce a dynamic sampling strategy: during the rollout phase, we calculate sample advantage values in real-time and discard zero-advantage samples. A policy update round is only triggered when the accumulated number of non-zero advantage samples reaches the preset batch size.

\textbf{Entropy Collapse Optimization.} During RL training, as the policy gradually converges, the entropy of the model's output distribution rapidly decreases, causing generation results to become deterministic. This premature stabilization hinders effective exploration of the reasoning manifold, making it difficult to discover better reasoning paths and ultimately constraining the improvement of the performance upper bound. To address this issue, we explored multiple strategies to slow down the entropy decline and maintain the model's exploration capability.

(1) Clip-higher\citep{DAPO}. We increase the clipping upper bound of the importance sampling ratio from the standard value to a larger value as shown in equation \ref{equ:clip_higher}, allowing tokens with lower probability under the current policy relative to the old policy to undergo greater probability increases, thereby enhancing exploration of low-frequency but potentially valuable reasoning paths. In our experiments, we adopt the officially recommended values: $\varepsilon=0.2$ and $\varepsilon_{high}=0.28$.
\vspace{-0.3cm}
\begin{equation}
\mathcal{J}_{\text{GRPO}}^{1}(\theta) = \mathbb{E}_{x \sim \mathcal{D}, \{y_i\}_{i=1}^G \sim \pi_{\theta_{\text{old}}}(\cdot|x)} \left[ \frac{1}{G} \sum_{i=1}^{G} \frac{1}{|y_i|} \sum_{t=1}^{|y_i|} \min\left(w_{i,t}(\theta)\widehat{A}_{i,t}, \text{clip}(w_{i,t}(\theta), 1-\varepsilon, 1+\varepsilon_{high})\widehat{A}_{i,t}\right) \right]
\label{equ:clip_higher}
\end{equation}

(2) On-policy. Traditional off-policy methods perform multi-epoch training on the same batch of experience data, using importance sampling constraints to prevent excessive policy deviation. However, this conservative update mechanism limits the model's exploration capability. To address this, we adopt a strictly on-policy strategy: each batch of sampled data ($batch\_size * group\_size * sample\_num$) is discarded after being used for only one gradient update, and the next update uses freshly sampled data from the new policy.

(3) Explicit Entropy Regularization. From the perspective of entropy preservation, we introduce an explicit entropy regularization term into the original GRPO loss function, where $\alpha$ is a balancing coefficient controlling the strength of entropy regularization. This term encourages the model to maintain randomness in its output distribution, directly counteracting entropy decay. Under this setup, we maintain the off-policy training mode (multi-epoch data reuse) and preserve exploration capability through an auxiliary loss term rather than through sampling strategy.

\begin{equation}
\begin{split}
\mathcal{J}_{\text{GRPO}}^{2}(\theta) = &\mathbb{E}_{x \sim \mathcal{D}, \{y_i\}_{i=1}^G \sim \pi_{\theta_{\text{old}}}(\cdot|x)} \left[ \frac{1}{G} \sum_{i=1}^{G} \frac{1}{|y_i|} \sum_{t=1}^{|y_i|} \min\left(w_{i,t}(\theta)\widehat{A}_{i,t}, \text{clip}(w_{i,t}(\theta), 1-\varepsilon, 1+\varepsilon)\widehat{A}_{i,t}\right) \right] \\
&\quad +\alpha \cdot \mathbb{E}_{x \sim \mathcal{D}, \{o_i\}_{i=1}^G \sim \pi_\theta(\cdot|x)} \left[ -\frac{1}{G} \sum_{i=1}^{G} \log \pi_\theta(o_i|x) \right]
\end{split}
\label{equ:entropy}
\end{equation}

\begin{figure}[h]
\centering
\subfigure[entropy comparison]{
    \includegraphics[width=0.45\textwidth]{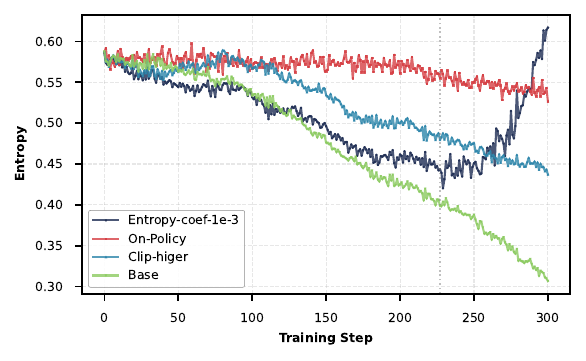}
}
\subfigure[reward comparison]{
\includegraphics[width=0.45\textwidth]{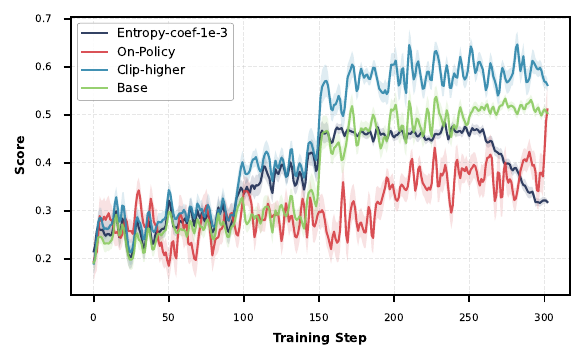}
}
\caption{Experimental comparison of three entropy collapse mitigation strategies: (1) clip-higher, (2) on-policy, and (3) off-policy with entropy loss.}
\label{fig:double}
\end{figure}

Fig \ref{fig:double} presents a comparison of several methods. The Clip-higher strategy exhibits optimal performance by effectively slowing the decline rate of policy entropy while maintaining its decreasing trend, achieving a favorable balance between exploration and exploitation, and ensuring training stability alongside significant performance improvements. The On-policy strategy stabilizes policy entropy, almost completely suppressing its decline. Although it maintains strong exploration capability, it simultaneously impedes effective policy convergence, resulting in limited performance gains. The off-policy + entropy loss strategy suffers from training instability issues. Although this method maintains a gradual entropy decline in early training, as training progresses and policy entropy drops below a certain threshold, the magnitude of entropy regularization loss gradually increases and surpasses the GRPO main loss to become the dominant term, triggering abnormal policy entropy increases that ultimately lead to training collapse. Furthermore, this method requires extensive hyperparameter tuning experiments to configure the entropy loss weighting coefficient, increasing application costs. In summary, the Clip-higher strategy achieves an optimal balance between exploration capability and convergence efficiency by moderately controlling the decay rate of policy entropy, making it the best choice for the scenario in this study.

\textbf{Importance Sampling Granularity Exploration.} 
The standard GRPO algorithm computes importance sampling ratios at the token level, i.e., independently calculating for each position t as shown in equation \ref{equ:ratio}. Inspired by GSPO\citep{GSPO} and GMPO\citep{GMPO}, we explored sentence-level importance sampling, which computes a unified importance weight for the entire output sequence based on sequence likelihood as shown in equation \ref{equ:GSPO}. This weight is shared by all tokens in the sequence, ensuring that the entire reasoning chain is consistently reinforced or suppressed as a whole.
\begin{equation}
\label{equ:GSPO}
w_i^{seq}(\theta) = \left(\frac{\pi_\theta(y_i|x)}{\pi_{\theta_{\text{old}}}(y_i|x)}\right)^{\frac{1}{|y_i|}} = \exp\left(\frac{1}{|y_i|}\sum_{t=1}^{|y_i|}\log\frac{\pi_\theta(y_{i,t}|x,y_{i,<t})}{\pi_{\theta_{\text{old}}}(y_{i,t}|x,y_{i,<t})}\right)
\end{equation}

\begin{figure}[h]
\centering
\subfigure[entropy comparison]{
    \includegraphics[width=0.45\textwidth]{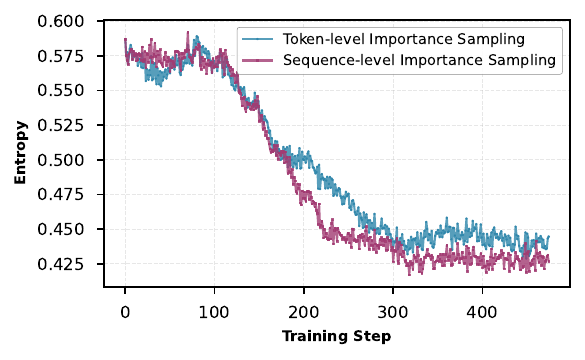}
}
\subfigure[reward comparison]{
    \includegraphics[width=0.45\textwidth]{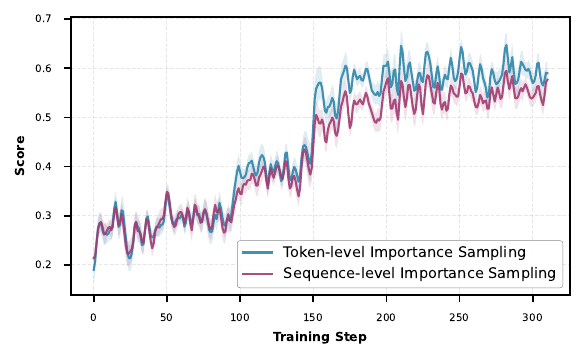}
}
\caption{Experimental comparison between token-level and sequence-level importance sampling.}
\label{fig:score_comparision}
\end{figure}
Fig~\ref{fig:score_comparision} presents the experimental results for different importance sampling granularities. Sentence-level importance sampling exhibits a more stable reward increase curve, but correspondingly, its policy entropy declines more rapidly. For tasks such as relevance judgment where reasoning patterns are relatively fixed, sentence-granularity probability adjustments more readily lead to premature solidification of model strategies, thereby constraining exploration capabilities. Consequently, its later-stage performance is inferior to that of token-level importance sampling.

\subsection{Evaluation: Comprehensive Relevance Benchmark}
\label{sec:3.5}
Following our detailed deconstruction of relevance judgment difficulty in Section~\ref{prel}, we find that existing e-commerce benchmarks, such as Shopping Queries~\cite{shoppingqueries}, fail to provide sufficient complexity for evaluation. Specifically, they lack the discriminative power to evaluate models on the more difficult judgment cases identified in our analysis. Consequently, they do not adequately challenge models in the critical competencies required for nuanced relevance judgment.

To address this gap, we developed the \textbf{R}ule-\textbf{A}ware benchmark with \textbf{I}mage for \textbf{R}elevance assessment (\textbf{RAIR}). Our objective is to provide a comprehensive, realistic, and challenging open-source benchmark for evaluating the relevance models proposed in this paper, and more broadly, for relevance tasks in the era of large models.

\subsubsection{Design Principles}
The design of RAIR must not only comprehensively evaluate a model's foundational judgment capabilities in realistic e-commerce scenarios, but must also specifically focus on the three core capability requirements identified in Section~\ref{prel}:

\begin{itemize}
    \item \textbf{Knowledge \& Reasoning Capability}: Relevance assessment often requires reasoning that combines world knowledge with inference, surpassing basic matching. A benchmark must test this ability. For instance, a model must access its knowledge base to recognize that "Capybara" is synonymous with "shui tun" to make an accurate judgment.
    
    \item \textbf{Multi-modal Comprehension Capability}: In real-world scenarios, visual modalities provide complementary, discriminative features that transcend textual representations. Item images are critical when textual features alone are insufficient. Therefore, evaluating VLM-based relevance modeling is essential for a forward-looking benchmark.
    
    \item \textbf{Rule Adherence Capability}: Relevance assessment is anchored in a sophisticated rule system crafted by domain experts. These rules transform subjective user preferences into objective, quantifiable protocols, acting as complementary specifications to common knowledge. For example, a model must adhere to the rule, "Without explicit user intent, a second-hand product is considered irrelevant," even if it otherwise matches the query.
\end{itemize}

\subsubsection{Benchmark Construction}
The construction of the RAIR benchmark was a multi-stage process based on data derived from Taobao user search logs, all of which underwent professional relevance annotation with quality assurance. The process was meticulously designed to create three distinct, complementary subsets, each targeting specific model capabilities.

\paragraph{General subset}
This subset aims to comprehensively assess foundational reasoning, knowledge, and attribute extraction capabilities using common e-commerce cases. A primary challenge was to mitigate the inherent biases of platform-specific traffic distributions, such as the dominance of fashion and cosmetics categories. Instead of naive uniform sampling, we implemented a two-step stratified sampling strategy. We first stratified queries by industry sector and then applied proportional sampling, followed by downsampling of dominant industries. This resulted in a balanced subset of 46,775 entries distributed across 14 industries, ensuring no single sector constitutes more than 15\% of the dataset.

\paragraph{Long-Tail Hard subset}
This subset was designed to provide challenging data to differentiate the capabilities of advanced LLMs. We focused on mining hard cases, which we categorize into explicit demands (\textbf{Multi-Attribute, MA}) and implicit demands. The implicit demands include four distinct types: \textbf{Negation (NE)}, \textbf{Alternatives (AL)}, \textbf{Knowledge-dependent (KD)}, and \textbf{Reasoning-dependent (RD)}. The mining process was tailored for these types. NE and AL queries were filtered using predefined keywords and regular expressions. KD and RD queries, which often appear as long-tail queries, were mined by first undersampling low-frequency logs and then using a powerful LLM (Qwen3-235B) with a self-consistency filter (requiring $>$ 4 identical results across 8 inferences). MR queries were identified using an NER model on length-filtered queries to find those with more than 5 attributes. To ensure the final set was genuinely difficult, all candidates were filtered through a baseline relevance model; we retained only those samples where the model's prediction was correct five or fewer times across eight inferences, indicating high model uncertainty.

\paragraph{Visual Salience subset}
This subset isolates cases where the main product image provides crucial, fine-grained visual information (e.g., color, style) that is indispensable for a correct judgment. We devised a multi-stage pipeline, starting with a keyword-based retrieval using a comprehensive visual taxonomy. To ensure visual necessity and mitigate noise from false recalls, we applied two critical filters: first, a predefined blacklist, and second, retaining only query-ad pairs where the triggering visual keyword *did not* appear in the product's textual description. An LLM was then used to verify that the remaining keywords genuinely referred to visual properties. Finally, to identify inherently difficult samples, we trained a naive multimodal relevance model and retained candidates that exhibited high prediction uncertainty (a consistency of less than four in a majority vote across eight predictions).

\paragraph{Rule Checklist Generation}
To bridge the gap between academic evaluation and real-world industrial scenarios, we augmented each sample with a checklist of the specific rules governing its judgment. This annotation was generated using an LLM in a "reverse-engineering" process. To manage the prompt's context length and improve accuracy, we stratified samples by relevance label ($y$) and industry ($c$) and provided the LLM with only the relevant subset of rules ($R'_{y,c}$). To ensure high fidelity, we performed $k$ independent inferences for each sample and applied a robustness filter: a rule $r$ was only added to the final checklist $R_{(Q,I)}$ if it was identified in at least two of the $k$ inferences. This checklist is intended strictly for evaluation.

\subsubsection{Data Statistics}
The RAIR benchmark encompasses a total of 63,601 samples, drawn from 14 real-world e-commerce industries. It is composed of the three distinct subsets established during our construction process: the \textbf{General subset} (46,775 samples), the \textbf{Long-Tail Hard subset} (10,931 samples), and the \textbf{Visual Salience subset} (5,895 samples). A detailed breakdown of the five challenging query intents within the Long-Tail Hard subset is presented in Table~\ref{tab:query_freq}. 

For each entry in the dataset, we provide the query, item title, item details, SKU information, and an anonymized item image. Each entry is accompanied by its manual ground truth annotation according to our four-level relevance scale (L1-L4). The comprehensive distribution of these labels across the dataset is presented in Table~\ref{tab:label_fre}. Additionally, to facilitate subsequent model development and detailed error attribution, each data instance is augmented with the corresponding rule identifiers (our \textit{Rule Checklist}) that serve as the basis for its relevance judgment.

\begin{table}
  \caption{Distribution of challenging query intentions within the long-tail hard subset}
  \centering
  \label{tab:query_freq}
  \begin{tabular}{cccc}
    \toprule
    Query Type & Query Intent & Num & Frequency\\
    \midrule
    Explicit demand & Multi-attribute & 5071 & 46.4\% \\
    \midrule
    \multirow{5}{*}{Implicit demand} & Negation & 1073 & 9.8\% \\
    & Alternatives & 213 & 1.9\% \\
    & Reasoning dependent & 2436 & 22.3\% \\
    & Knowledge dependent & 2138 & 20.0\% \\
    \bottomrule
  \end{tabular}
\end{table}

\begin{table}[htbp]
  \centering  
  \caption{Distribution of ground truth labels in RAIR}
  \label{tab:label_fre}
  \begin{tabularx}{\columnwidth}{>{\centering\arraybackslash}X >{\centering\arraybackslash}X >{\centering\arraybackslash}X >{\centering\arraybackslash}X}  
    \toprule
    Relevance  & Ground Truth & Num & Frequency \\
    \midrule
    \multirow{2}{*}{Irrelevant} & L1 & 2985 & 4.7\% \\
    & L2 & 13443 & 21.1\% \\
    \midrule
    \multirow{2}{*}{Relevant} & L3 & 3034 & 4.7\% \\
    & L4 & 44139 & 69.4\% \\ 
    \bottomrule  
  \end{tabularx}
\end{table}

\section{Experiment}
In this section, we conduct a systematic validation and analysis of the proposed methods on the RAIR benchmark. Our experiments are divided into two parts: first, through quantitative metrics on benchmarks, we validate the following two core questions: (1) the magnitude of performance improvement brought to the model by fine tuning; (2) the performance advantages demonstrated by the proposed LORE model across different data distribution, including general samples, hard samples, and visually salient samples. Second, we employ case studies to illustrate the new capabilities that our methods confer upon the model and the types of problems they enable it to solve. 

\subsection{Settings}
\subsubsection{Metrics}
To ensure the stability and reproducibility of evaluation results, all trained models employ a greedy decoding strategy for single-pass inference, without introducing sampling randomness. We employed a range of metrics to evaluate the model's performance. The dataset provides annotations from L1 to L4, thus we first introduced the four-class accuracy metric (acc@4) as follows:
\begin{equation}
\label{equ:3}
\text{acc@4} = \frac{1}{N}\sum_{i=1}^{N}\mathbb{I}(\hat{y}_i = y_i)
\end{equation}

Given the nature of the relevance task, where L1 and L2 can be considered as irrelvant and L3 and L4 as relevant, we introduced the binary classification accuracy metric (acc@2) as follows: 
\begin{equation}
\label{equ:4}
\text{acc@2} = \frac{1}{N}\sum_{i=1}^{N}\mathbb{I}(\hat{y}_i,y_i \in (L1,L2) | \hat{y}_i,y_i \in (L3,L4))
\end{equation}
Additionally, to mitigate the impact of the imbalance in data label distribution, we introduce the macro F1 score as follows, which is calculated as the average of F1 scores across all classes.
\begin{equation}
\label{equ:4}
F1\text{-}score_i = 2\frac{\text{Recall}_i \times \text{Precision}_i}{\text{Recall}_i + \text{Precision}_i} 
\end{equation}
\begin{equation}
\label{equ:4}
macro-F1 = \frac{1}{N}\sum_{i=1}^{N}F1-score_i
\end{equation}

\subsubsection{Baseline}
As comparison baselines, we select the following two categories of models: (1) Top-tier large language models (open-source and closed-source): All models are evaluated using zero-shot direct prompting with a unified prompt template. The prompts comprise the complete relevance task definition and discrimination rules. The inference configuration follows the best practices reported in the official documentation of each model: employing single-pass inference with temperature parameters set to the officially recommended optimal values for each model (e.g., 0.7 for GPT-5, 0.6 for Llama-3, etc.), ensuring that each model is compared under its optimal operating conditions. (2) Vanilla SFT Model: This model is trained using only labels, without incorporating CoT modeling or the RL phase, serving to validate the joint contribution of CoT modeling and reinforcement learning optimization in our method.

\subsection{Main Result}

Table \ref{tab:main_results} and Table \ref{tab:visual_results} present the main experimental results. The analytical conclusions of this study are summarized as follows: 

\textbf{Domain-specific fine-tuning yields substantial performance gains. }Across both the General subset and Hard subset, supervised fine-tuned models demonstrate significant improvements over their base models, with performance even surpassing large-scale closed-source state-of-the-art (SOTA) models such as GPT-5. This enhancement is particularly pronounced on the General subset: LORE achieves relative gains of 8.8\% and 29.1\% over GPT-5 in Acc@2 and Macro-F1 metrics, respectively. This improvement is primarily attributed to the model's effective acquisition of business rule modeling capabilities and domain-specific knowledge in e-commerce.

\begin{table}[t]
\centering
\renewcommand{\arraystretch}{1.2}
\caption{Main results on the general subset and long-tail hard subset. The best score in each column is in \textbf{bold}, and the second-best is \underline{underlined}.}
\label{tab:main_results}
\resizebox{\textwidth}{!}{%
\begin{tabular}{lcccccc}
\toprule
\multirow{2}{*}{Model} & \multicolumn{3}{c}{General subset} & \multicolumn{3}{c}{Longtail Hard subset} \\
\cmidrule(lr){2-4} \cmidrule(lr){5-7}
& acc@2 & acc@4 & macro-F1 & acc@2 & acc@4 & macro-F1 \\
\midrule
\multicolumn{7}{l}{\textit{prompting-based models}} \\
\hline
Qwen2.5-7B(base)\citep{qwen25} & 0.775 & 0.689 & 0.395 & 0.531 & 0.382 & 0.312 \\
Qwen3-30B-Instruct\citep{Qwen3} & 0.763 & 0.683 & 0.391 & 0.587 & 0.431 & 0.333 \\
Qwen3-235B-Instruct\citep{Qwen3} & 0.830 & 0.676 & 0.417 & 0.609 & 0.381 & 0.359 \\
Llama3.1-8B-Instruct\citep{llama3} & 0.788 & 0.416 & 0.224 & 0.525 & 0.248 & 0.200 \\
Llama3.1-70B-Instruct\citep{llama3} & 0.774 & 0.668 & 0.369 & 0.547 & 0.378 & 0.295 \\
Qwen3-4B-Thinking\citep{Qwen3} & 0.805 & 0.720 & 0.453 & 0.584 & 0.419 & 0.357 \\
Qwen3-30B-Thinking\citep{Qwen3} & 0.784 & 0.718 & 0.457 & 0.582 & 0.448 & 0.362 \\
Qwen3-235B-Thinking\citep{Qwen3} & 0.779 & 0.702 & 0.470 & 0.585 & 0.451 & 0.367 \\
Gemini 2.5 Pro\citep{gemini2.5} & 0.795 & 0.701 & 0.483 & 0.627 & 0.481 & 0.392 \\
GPT-5 & 0.845 & 0.714 & 0.433 & \underline{0.681} & 0.435 & 0.407 \\
\midrule
vanilla SFT & \underline{0.929} & \underline{0.891} & \underline{0.722} & 0.671 & \underline{0.542} & \underline{0.413} \\
\midrule
LORE & \textbf{0.933} & \textbf{0.897} & \textbf{0.724} & \textbf{0.715} & \textbf{0.582} & \textbf{0.460} \\
\bottomrule
\end{tabular}%
}
\end{table}

\begin{table}[t]
\centering
\renewcommand{\arraystretch}{1.2}
\caption{Results on the Visual salience subset. The best score in each column is in \textbf{bold}, and the second-best is \underline{underlined}.}
\label{tab:visual_results}
\begin{tabular}{lccc}
\toprule
\multirow{2}{*}{Model} & \multicolumn{3}{c}{Visual salience subset}\\
\cmidrule(lr){2-4}
& acc@2 & acc@4 & macro-F1 \\
\midrule
\multicolumn{4}{l}{\textit{prompting-based models}} \\
\hline
Qwen2.5-VL-7B-Instruct\citep{qwen2.5vl} & 0.535 & 0.285 & 0.230 \\
Qwen2.5-VL-32B-Instruct\citep{qwen2.5vl} & 0.647 & 0.467 & 0.339 \\
Qwen2.5-VL-72B-Instruct\citep{qwen2.5vl} & 0.608 & 0.420 & 0.267 \\
Gemini 2.5 Pro\citep{gemini2.5} & 0.670 & 0.561 & 0.377 \\

GPT-5 & \underline{0.682} & 0.508 & 0.369 \\
\midrule
    vanilla SFT & 0.638 & \underline{0.574} & \underline{0.378} \\
\midrule
LORE & \textbf{0.698} & \textbf{0.627} & \textbf{0.426} \\
\bottomrule
\end{tabular}
\end{table}

\textbf{Multi-dimensional CoT modeling yields marginal improvements on high-performance baselines.} On the General subset, the vanilla-SFT model achieves an Acc@2 of 0.929 and a Macro-F1 of 0.722, demonstrating performance significantly superior to advanced models such as GPT-5. Compared to vanilla-SFT, the LORE model further improves Acc@2 by 0.4 percentage points. This indicates that although vanilla-SFT has already achieved sufficient fitting for in-distribution routine data, LORE can still obtain marginal performance gains on top of this high-performance baseline.

\textbf{The injection of reasoning and knowledge capabilities yields substantial gains for LORE on hard samples.} On the Long-Tail Hard subset, vanilla-SFT demonstrates a smaller performance improvement relative to the base model compared to its performance on the General subset, and fails to establish a significant performance gap with models such as GPT-5 (0.671 vs 0.681 on acc@2). This is because for hard samples, the key factors constraining model performance are not discriminative rules, but rather reasoning capabilities and knowledge reserves. In contrast, LORE achieves a significant breakthrough: attaining an Acc@2 of 0.715 (a relative improvement of 4.4\%) and a Macro-F1 of 0.460 (a relative improvement of 5.3\%), surpassing large language models with strong reasoning capabilities such as GPT-5. These experimental results provide compelling evidence that knowledge distillation through teacher models and the injection of synthetic CoT reasoning can effectively enhance the model's discriminative capability for hard samples.


\textbf{Relevance-guided enhanced caption significantly improves multimodal performance.} As shown in Table \ref{tab:visual_results}, on samples requiring image information for discrimination, vanilla-SFT, which lacks visual information, does not demonstrate a significant advantage over a series of advanced Vision-Language Models (VLMs). In contrast, LORE, which employs a visual relevance enhancement strategy, achieves a breakthrough: attaining a Macro-F1 of 0.426, representing a relative improvement of 4.8\% compared to strong baseline models such as GPT-5. These results indicate that CoT modeling incorporating multimodal information can effectively enhance the model's cross-modal understanding and reasoning capabilities.

\subsection{Case Study}
\begin{figure}[t]
  \centering
  \includegraphics[width=0.95\textwidth]{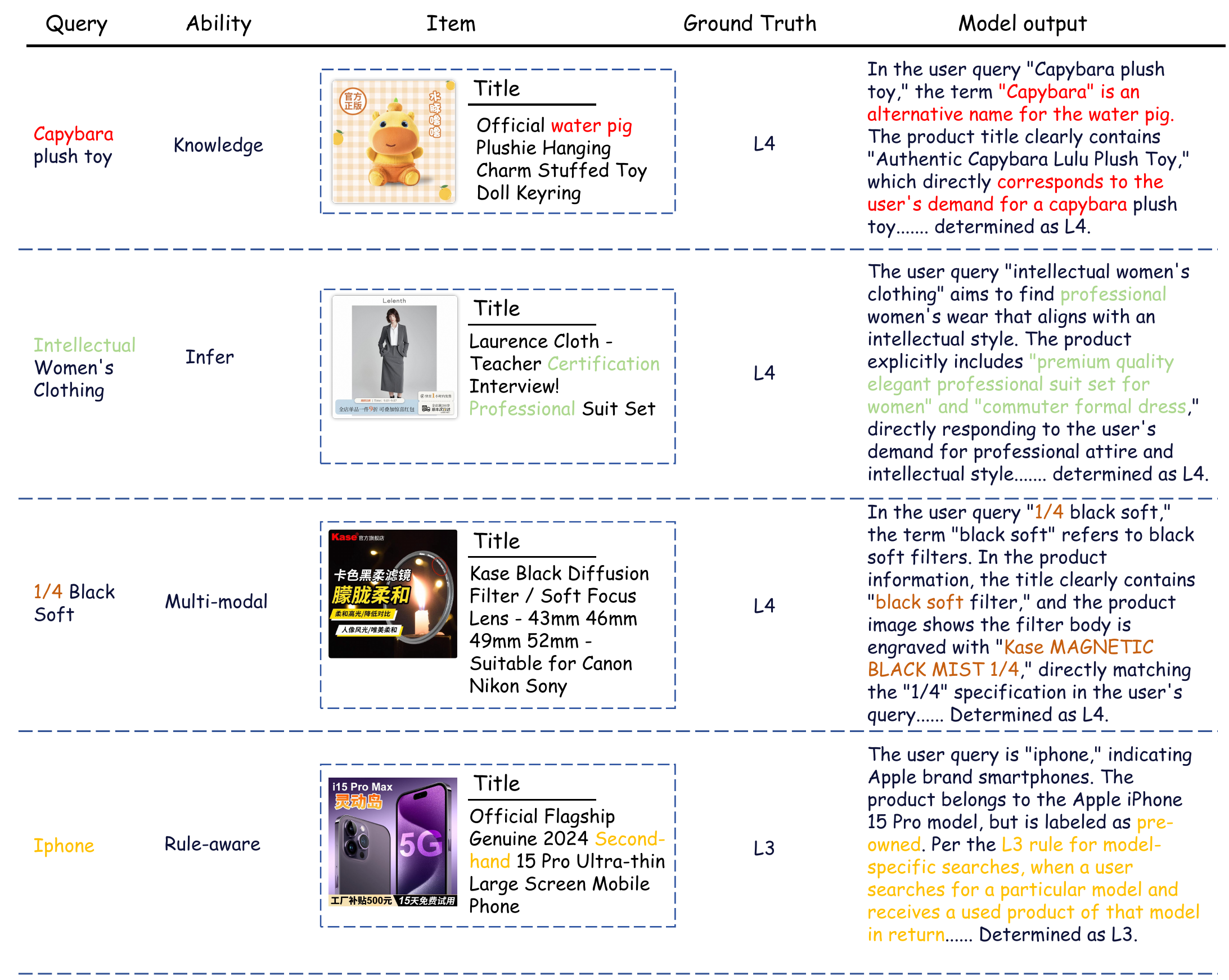}
  \caption{Example analysis of capabilities required for relevance evaluation.}
  \label{fig:case}
\end{figure}
From a qualitative perspective, we demonstrate the model's improvements across multiple capability dimensions through representative cases, as shown in Fig \ref{fig:case}:

\textbf{Knowledge.}
In Case 1, the query term is "Capybara" and the product title is "water pig plush toy." After distillation from the teacher model, the model has internalized sufficient world knowledge to recognize that "Capybara" is an alternative name for "water pig," thereby correctly determining their relevance.

\textbf{Reasoning.}
In Case 2, the query is "intellectual women's clothing", while the product information contains no directly matchable literal keywords. Through reasoning capability, the model analyzes that: the product "commuter formal dress" aligns with the user's implicit demand for professional attire and intellectual style; the color (gray), applicable age range (18-24 years old), and design elements (suit set, formal dress) in the product details all conform to the typical requirements of the intellectual demographic for simple and appropriate attire, leading to a correct judgment.

\textbf{Multimodal Understanding.}
Case 3 represents a typical scenario of joint text-image discrimination. For the query "1/4 Black Soft," the model matches the "Black Soft" attribute in the product title while simultaneously locating the key text "Kase MAGNETIC BLACK MIST 1/4" engraved in a small area of the product image, achieving coordinated reasoning between textual and visual information. This benefits from the relevance-oriented caption generation strategy (Section 3.3), enabling the model to extract task-critical information from images.

\textbf{Rule Compliance.}
In Case 4, the attribute matching between query and product is not inherently complex, but the sample belongs to a special case stipulated by rules: "searching for normal products but returning second-hand products should be judged as L3 (weakly relevant)." After rule-aware CoT modeling (Section 3.3), the model has acquired sufficient rule compliance capability to accurately identify and apply such business rules.

\section{Application}
\begin{figure}[htbp]
  \centering
  \includegraphics[width=0.95\textwidth]{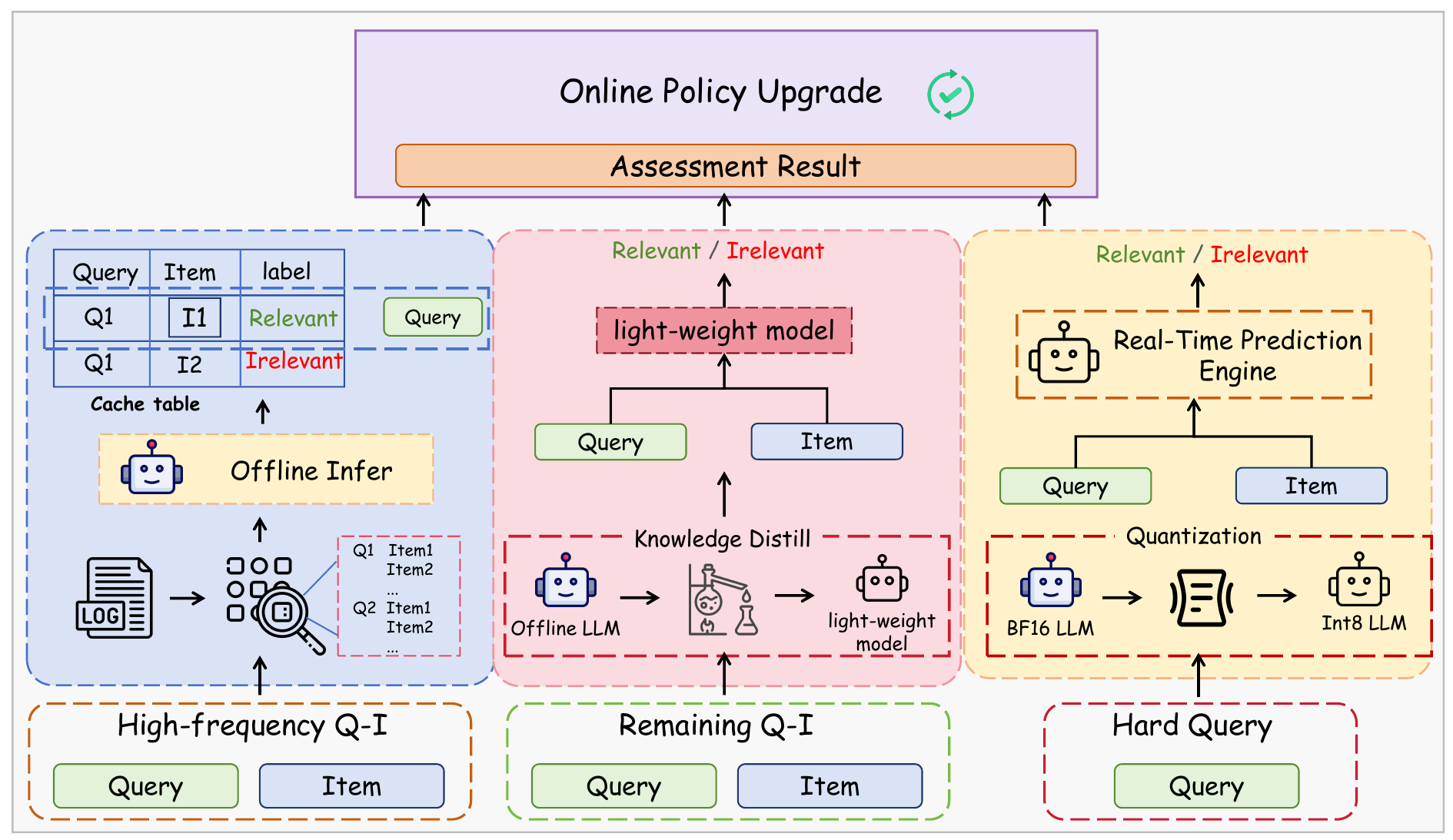}
  \caption{Online application approach of LORE based on query frequency segmentation.}
  \label{fig:application}
  \vspace{-0.4cm}
\end{figure}

Based on the robust model performance, we design a complete solution encompassing both Model \& Serving Update and System \& Strategy Update, significantly enhancing the overall performance of end-to-end relevance discrimination, as illustrated in Fig~\ref{fig:application}. For the former, based on query stratification, we design three strategies to leverage model capabilities to improve online performance.
For the System \& Strategy Update, we upgrade the online strategy and infrastructure leveraging LLM capabilities. The experimental results demonstrate that our solution achieves a remarkable cumulative gain of 27\% in GoodRate, as presented in Table \ref{tab:online}.

\begin{table}[h]
\centering
\renewcommand{\arraystretch}{1.2}
\caption{Performance gains from online deployment of LORE. $^{\ast}$denotes estimated improvements derived from offline evaluation; 
actual online deployment results are pending.}
\label{tab:online}
\begin{tabular}{llc}
\toprule
\textbf{Category} & \textbf{Approach} & \textbf{GoodRate} \\
\midrule
\multicolumn{2}{c}{Overall} & +27.0\% \\
\midrule
\multicolumn{2}{c}{System \& Strategy Update} & +12.7\% \\
\midrule
\multirow{3}{*}{Model \& Serving Update} & Cache Deployment & +4.8\% \\
\cline{2-3}
 & Knowledge Distillation & +9.5\% \\
\cline{2-3}
 & Real-time Inference Strategy$^{\ast}$ & +0.9\% \\
\bottomrule
\end{tabular}
\end{table}

(1) Cache Deployment: for high-frequency query-item pairs (approximately 30\% of total traffic), which occur frequently in e-commerce systems and hold significant commercial value, we aim to directly leverage LLM capabilities on these samples. To address the latency bottleneck imposed by real-time computation, we adopted a hybrid deployment strategy combining offline pre-computation with online caching. Specifically, we sampled a batch of high-frequency query-item pairs from search logs. For each query, a variable number of items were sampled to construct query-item pairs, which were then processed via offline inference using the LLM.

(2) Knowledge Distillation: for medium-frequency query-item pairs (approximately 65\% of total traffic), which constitute a substantial proportion. Pre-computing and caching all query-item discrimination results is neither feasible nor cost-effective. Therefore, we maintained the original "coarse ranking-fine ranking" processing pipeline. Meanwhile, the LLM was leveraged to generate large-scale, high-quality training data, systematically enhancing the discrimination capabilities of both online coarse ranking and fine ranking models through knowledge distillation techniques.through knowledge distillation techniques\citep{ELLM}.

(3) Real-time Inference: for hard queries(approximately 5\% of total traffic), which typically involve complex knowledge reasoning scenarios, online models struggle to make accurate predictions due to both parameter limitations and insufficient training from sparse data. While caching approaches prove ineffective given the sparsity of these pairs, these challenging cases present an opportunity for direct LLM intervention. To address this, we are actively working on developing an online real-time LLM inference capability. First, we plan to train a lightweight intent recognition model to identify these challenging queries. To address the aforementioned latency constraints, we will apply model quantization to accelerate inference speed and initially deploy this solution to a the subset of traffic. Experimental results on offline hard set demonstrate a remarkable 14\% improvement in GoodRate, and we anticipate achieving approximately 0.89\% gain in overall metrics once this capability is fully deployed to production traffic.

(4) System \& Strategy Update: We fully leverage LLM capabilities to upgrade both online strategies and system infrastructure. From the strategic perspective, we incorporate the more reliable query-item relevance scores produced by the enhanced model into the downstream ranking module and derive the optimal trade-off via Pareto optimization. From the infrastructure perspective, supported by the model's robust performance guarantees, we systematically retire or substitute heuristic rules previously employed in the pipeline. The integrated deployment of both dimensions demonstrates substantial gains.

\section{Discussion}
\label{sec:discuss}

\subsection{Naive teacher CoT distillation results in negative effects.}
\label{sec:6.1}
\begin{table}[h]
\centering
\caption{Performance comparison of single-pass and multi-pass inference between vanilla SFT and cold-start model with CoT distillation.}
\renewcommand{\arraystretch}{1.2}  
\setlength{\tabcolsep}{1.2\tabcolsep}
\begin{tabular}{lccc}
\toprule
 Models & pass@1 & pass@8 \\
\midrule
vanila SFT & 0.929 &  0.937\\
\hline
cold-start & 0.887(-4.2\%) & 0.964(+2.7\%)\\
\bottomrule
\end{tabular}
\label{tab:naive_SFT}
\end{table}

In our CoT data distillation experiments, we observed that distilling synthetic CoT data paradoxically degrades model performance compared to vanilla SFT. As shown in Table~\ref{tab:naive_SFT}, the distilled model exhibits a 4.2\% decline in pass@1 metrics relative to the Naive-SFT baseline. This degradation stems from the training-inference distribution shift: during training, the model conditions on high-quality contexts generated by the teacher, whereas during inference, it relies on its own autoregressive outputs. This discrepancy triggers error accumulation through the reasoning chain, ultimately leading to incorrect predictions. Notably, however, the distilled models demonstrate significant improvements in pass@8 metrics (2.7\% increase), indicating that distillation enhances the model's capability upper bound by expanding answer space diversity and instilling multi-step reasoning proficiency.

\subsection{Long CoT is not necessary for better performance.}
\label{sec:6.2}
\begin{figure}[htbp]
  \centering
  \includegraphics[width=1\textwidth]{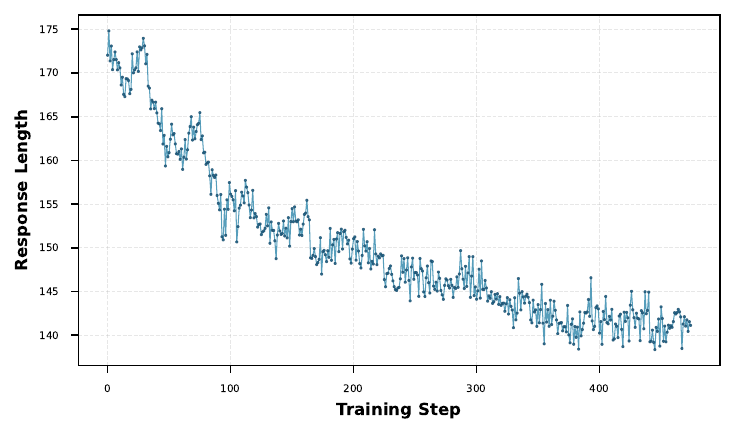}
  \caption{Response length variation during RL training.}
  \label{fig:length}
\end{figure}
Studies such as DeepSeek-R1 have observed the emergent phenomenon of long CoT and the model's "Aha Moment." We similarly conducted a tracking analysis of the dynamic changes in model output length during the reinforcement learning process, with results presented in Fig \ref{fig:length}. The experimental findings reveal that as model capability continues to improve, the output length exhibits an overall decreasing trend and eventually stabilizes. This phenomenon suggests that long CoT are not a necessary condition for model capability enhancement, but rather an accompanying phenomenon during the training process. Further analysis indicates that different task types exhibit significant variations in their requirements for reasoning chain length: tasks such as mathematical computation and code generation typically necessitate longer step-by-step reasoning processes. For the relevance task, accurately identifying query intent and item attributes, and reasoning based on discrimination rules, constitutes an intuitive and clear chain of thought. Redundant reasoning processes beyond this do not yield performance improvements.

\subsection{Multimodal Modeling: VLM or Two-Stage LLM?}
A more intuitive approach to modeling multi-modal capabilities for relevance tasks is to directly leverage VLMs, as they inherently possess image understanding capabilities without requiring additional caption synthesis. We conducted exploratory experiments on this approach: we employs the teacher model Qwen2.5-VL-72B to sequentially complete two steps—knowledge and reasoning CoT generation, followed by rule-aware CoT generation—thereby synthesizing multi-dimensional CoT data. Subsequently, using the same two-stage training configuration (SFT and RL), we trained the Qwen2.5-VL-7B model, which has a comparable parameter to the LLM base model employed in this study.
\begin{table}[h]
\centering
\renewcommand{\arraystretch}{1.2}
\caption{Performance comparison of VLM-based and LLM-based methods on the general subset and long-tail hard subset. The best score in each column is in \textbf{bold}, and the second-best is \underline{underlined}.}
\label{tab:vlm_general}
\begin{tabular}{lcccccc}
\toprule
\multirow{2}{*}{Model} & \multicolumn{3}{c}{General subset} & \multicolumn{3}{c}{Longtail Hard subset} \\
\cmidrule(lr){2-4} \cmidrule(lr){5-7}
& acc@2 & acc@4 & macro-F1 & acc@2 & acc@4 & macro-F1 \\
\midrule
VLM-base & 0.912 & 0.878 & 0.670 & 0.692 & 0.577 & 0.413 \\
LLM-base & 0.933 & 0.897 & 0.724 & 0.715 & 0.582 & 0.460 \\
\bottomrule
\end{tabular}
\end{table}

\begin{table}[h]
\centering
\renewcommand{\arraystretch}{1.2}
\caption{Performance comparison of VLM-based and LLM-based methods on the Visual salience subset. The best score in each column is in \textbf{bold}, and the second-best is \underline{underlined}.}
\label{tab:vlm_visual}
\begin{tabular}{lccc}
\toprule
\multirow{2}{*}{Model} & \multicolumn{3}{c}{Visual Salient subset.}\\
\cmidrule(lr){2-4}
& acc@2 & acc@4 & macro-F1 \\
\midrule
VLM-base  & 0.703 & 0.645 & 0.526 \\
LLM-base  & 0.698 & 0.627 & 0.426\\
\bottomrule
\end{tabular}
\end{table} 

Tables \ref{tab:vlm_general} and \ref{tab:vlm_visual} present a performance comparison between trained models based on VLM and based on LLM. On both General and hard datasets, the LLM-based method significantly outperforms the VLM-based method, particularly demonstrating a clear advantage in macro-F1 metrics. This indicates that with equivalent CoT capability injection, LLMs possess stronger reasoning and knowledge capabilities compared to VLMs. In contrast, VLMs perform better on the visual salience subset, which aligns with expectations, as VLMs do not require generating intermediate textual captions, thereby avoiding information loss during the image-to-text conversion process.

\section{Evolutionary Trajectory}
\begin{figure}[htbp]
  \centering
\includegraphics[width=1\textwidth]{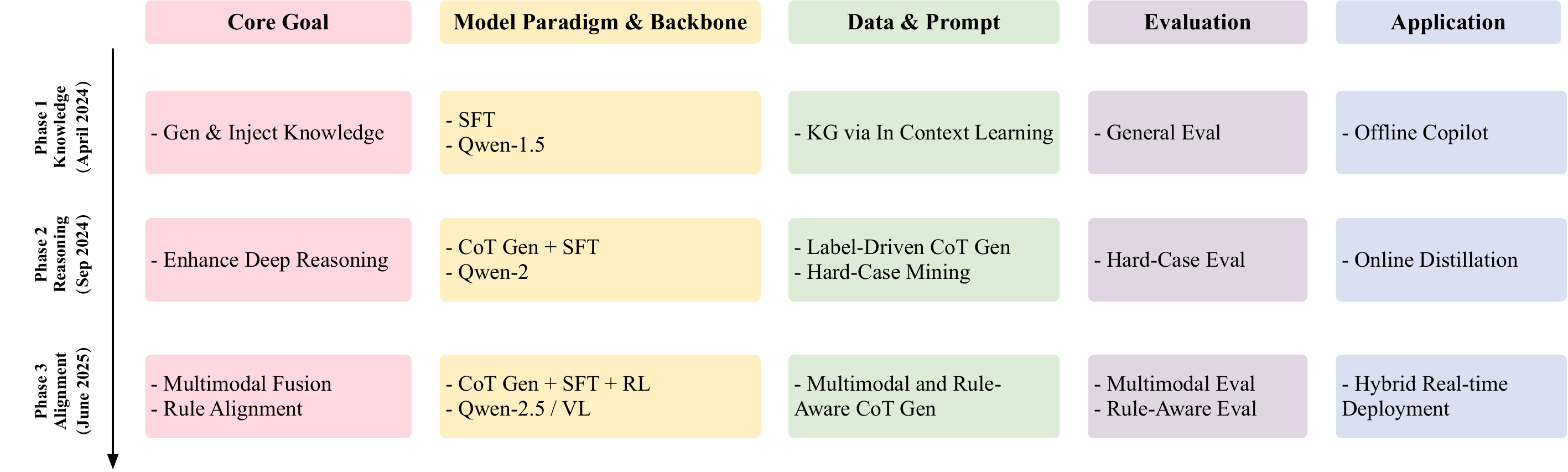}
  \caption{Three-phase evolution of LORE system across five key dimensions.}
  \label{fig:Trajectory}
\end{figure}

As shown in Fig~\ref{fig:Trajectory}, over the past three phases, our LORE model has undergone systematic evolution characterized by progressive refinements along five key dimensions: (1) core goals; (2) model paradigm and backbone; (3) data and prompt strategies; (4) evaluation frameworks; and (5) application scenarios. Together, these three phases trace a systematic trajectory of innovation that combines increasingly sophisticated model architectures, high-quality data construction pipelines, and practical deployment strategies to enhance search relevance capabilities in e-commerce scenarios. In this section, we outline this evolution across these core dimensions.

\subsection{LORE 1.0: Foundation Consolidation}
In this initial phase, we confronted the fundamental challenge of Query understanding in the e-commerce domain, particularly the accurate identification of entity mentions with non-standard expressions (e.g., unregistered terms or colloquialisms), which posed significant obstacles to robust product attribution understanding. To address this challenge, we constructed large-scale, high-quality e-commerce training corpora and employed In-Context Learning (ICL) to systematically inject domain knowledge into the model, thereby establishing a solid foundation for e-commerce semantic comprehension.

\textbf{Model Paradigm \& Backbone}: We adopted a Supervised Fine-Tuning (SFT) paradigm based on the Qwen-1.5 model as our backbone architecture. This approach enabled effective transfer of general language understanding capabilities to domain-specific e-commerce scenarios.

\textbf{Data \& Prompt}: Our methodology leveraged knowledge graphs combined with In-Context Learning, where domain knowledge was injected through carefully designed contextual prompts. This strategy allowed the model to assimilate e-commerce-specific semantic patterns and entity relationships through upper-context learning mechanisms.

\textbf{Evaluation}: We employed standard benchmark protocols alongside general relevance assessment metrics to comprehensively evaluate model performance across various query understanding tasks, ensuring both accuracy and generalizability.

\textbf{Application}: The trained model was deployed as a high-precision offline judger for label annotation tasks. This application successfully established an efficient annotation assistance system and badcase detection mechanism, significantly enhancing data quality and operational efficiency in downstream workflows.

\subsection{LORE 2.0: Deep reasoning}
Building upon the foundational query understanding capabilities, this phase tackled more sophisticated challenges in complex Query semantic recognition (including existence, implicit meanings, and negations) and deep semantic reasoning within Path construction. Simple instruction-following proved insufficient for these multi-step reasoning and judgment tasks that demanded enhanced logical inference capabilities. To address these challenges, we introduced Chain-of-Thought (CoT) technology. Rather than directly requesting judgment outputs, we strengthened the model's ability to "identify complex entities and semantics, decompose hierarchical relationships, and ultimately complete Path comparisons" through comprehensive reasoning chains. This approach significantly enhanced the accuracy of complex Query discrimination. Additionally, we established a data synthesis engine based on difficult case mining, enabling iterative feedback optimization.

\textbf{Model Paradigm \& Backbone}: We synthesized CoT data for attribute extraction and matching tasks, and trained the model using the SFT paradigm with Qwen-2 as the backbone architecture. This approach enabled the model to generate intermediate reasoning steps while maintaining task-specific optimization through supervised fine-tuning.

\textbf{Data \& Prompt}: Our methodology employed label-driven CoT generation combined with hardcase mining. By integrating label-constrained CoT synthesis with difficult sample excavation, we enhanced the model's reasoning capabilities through structured multi-step inference patterns and challenging edge cases.

\textbf{Evaluation}: We implemented hardcase-focused evaluation protocols specifically designed to assess performance on challenging semantic scenarios, ensuring robust handling of complex reasoning requirements and edge cases.

\textbf{Application}: The innovations in this phase enabled online distillation deployment, where the CoT-based multi-dimensional knowledge system was successfully distilled and transferred to smaller models. This achievement marked a significant milestone in scaling deep reasoning capabilities to production environments with improved inference efficiency, realizing substantial performance improvements in online A/B testing scenarios.

\subsection{LORE 3.0: Rule Adherence and Multi-modality}
Entering the deep-water zone, this phase confronted two paramount challenges: first, the high dependence on visual information in product attribution understanding (where user-uploaded images often contain insufficient textual details); and second, during the Path selection process, judgment results must strictly conform to formats that encompass numerous fine-grained rules and detailed business standards that align with observable attributes.

To address these challenges, we pursued a comprehensive technical strategy that integrated multimodal understanding capabilities with rule-aware reasoning mechanisms, enabling the model to process visual information while maintaining strict adherence to business constraints.

\textbf{Model Paradigm \& Backbone}: We employed a hybrid paradigm of CoT Generation + SFT + Reinforcement Learning (RL) built upon both Qwen-2.5 LLM for text reasoning and Qwen-2.5 VL for multimodal understanding. This dual-model architecture enabled seamless integration of visual and textual reasoning pathways, breaking through single-modality limitations.

\textbf{Data \& Prompt}: Our methodology leveraged multimodal and rule-aware CoT generation, combining image-text paired understanding with standard-oriented prompt engineering. First, business rules and exemplar cases were embedded into CoT reasoning chains to establish rule comprehension logic. Subsequently, reinforcement learning was introduced to significantly enhance the model's generalized reasoning capabilities for applying active rules to Path judgments under complex attribute scenarios.

\textbf{Evaluation}: We constructed a challenging benchmark, RAIR (Relevance Assessment with Image and Rules), designed to comprehensively evaluate the model across multiple dimensions: foundational capabilities, hardcase handling, multimodal understanding, and rule compliance. This holistic assessment framework ensures robust validation of all critical competencies required for production deployment.

\textbf{Application}: This phase achieved breakthrough hybrid real-time deployment capabilities. Through efficient high-frequency caching combined with long-tail real-time inference mechanisms, we successfully realized direct large-model discrimination capabilities online for the first time, driving the system toward next-generation platforms for online relevance experience validation.

\section{Conclusion}
In this study, we introduce LORE, a systematic and end-to-end framework designed to comprehensively address the multifaceted challenges of e-commerce search relevance. Our work began with a foundational deconstruction of the relevance task, identifying three core capabilities essential for robust judgment: (1) Knowledge and Reasoning Integration, (2) Multi-modal Understanding, and (3) Complex Rule Adherence. This theoretical analysis provided clear and structured guidance for relevance modeling. Building on this foundation, we proposed a complete and practical blueprint that spans the entire lifecycle of a relevance model. This methodology encompasses four key contributions:

\textbf{Systematic preliminary explorations} to optimize crucial foundational components, including feature engineering, prompt design, and base model selection, thereby establishing a robust groundwork for subsequent stages.

\textbf{A sophisticated training paradigm}, featuring a progressive CoT synthesis for capability injection via Supervised Fine-Tuning (SFT), followed by a Reinforcement Learning (RL) stage to align the model with human preferences.

\textbf{A comprehensive evaluation benchmark}, RAIR, which was specifically constructed to rigorously assess model performance across the identified core capabilities, especially on challenging long-tail and visually-salient samples.

\textbf{A practical online application strategy}, where we successfully deployed LORE into a large-scale production environment. By employing a query frequency-stratified approach, we achieved a significant 10\% cumulative improvement in the online GoodRate metric.

Experimentally, LORE not only established a new state-of-the-art on our challenging offline benchmark, outperforming leading proprietary models, but also demonstrated its substantial real-world value through impressive online gains.
In conclusion, LORE provides more than just a high-performing model; it offers a complete, replicable blueprint—from foundational exploration and principled training to robust evaluation and effective online deployment—for developing and operationalizing advanced relevance systems in the e-commerce industry, while also providing valuable insights for post-training work in other vertical domains.

\section{Contributors}
\label{sec:contributor}
\textbf{Core Contributors}

\textbf{Algorithm:} Chenji Lu, Zhuo Chen, Hui Zhao, Zhiyuan Zeng, Ximing Zhang, Gang Zhao, Junjie Ren, Ruicong Xu, Haoran Li, Songyan Liu, Zhenyi Wang, Pengjie Wang, Jian Xu, Bo Zheng

\textbf{Engineering}: Qingtao Zeng, Huimin Yi, Siqi Luo, Kaiwen Wei, Chen Chen, Fei Wang, Xiaojie Zhang, Qifeng Li, Daisong Wang, Qing Xue, Jiawen Liao, Zexin Yan, Zhenyuan Lai, Peng Sun, Jiangwei Yu, Zhige Chen, Yang Wang, Rong Huang, Shihao Chen, Xiang Gao, Ju Huang, Weixun Wang, Zhendong Li, Feilei Du, Jiamang Wang

\textbf{Annotation}: Weihao Guo, Junjing Su, Nan Lin, Xinyao Zhou, Annotation Team
\newpage

\bibliographystyle{plainnat}
\bibliography{colm2024_conference}
\end{document}